\documentclass[traditabstract]{aa}

\usepackage{graphicx}
\usepackage{txfonts}
\usepackage{longtable}

\usepackage{epsf}
\usepackage{rotating}
\usepackage{graphics}

\def \MJ{M$_{\mathrm{Jup}}$}
\def \RJ{R$_{\mathrm{Jup}}$}

\def \kms{km\,s$^{-1}$}
\def \ms{m\,s$^{-1}$}
\def \1s{$1\,\sigma$}
\def \kid{$\chi^2$}
\def \t0{T$_0$}

\newcommand{\kepler}{\emph{Kepler}}
\newcommand{\corot}{\emph{CoRoT}}
\newcommand{\Mjup}{M$_\mathrm{Jup}$}
\newcommand{\Rjup}{R$_\mathrm{Jup}$}

\def\Msun{\hbox{$\mathrm{M}_{\odot}$}}            
\def\Rsun{\hbox{$\mathrm{R}_{\odot}$}}
\def\met{[Fe/H]}
\def\mr{$M_\star^{1/3}/R_\star$}

\newcommand{\teff}{\mbox{$T_{\rm eff}$}}
\newcommand{\logg}{\mbox{$\log g_*$}}
\newcommand{\vsini}{\mbox{$v \sin i_*$}}

\def \kd{KOI-200}
\def \kdb{{\kd}\,b}
\def \kh{KOI-889}
\def \khb{{\kh}\,b}

\begin{document}

   \title{KOI-200\,b and KOI-889\,b: two transiting exoplanets detected \\and characterized with \emph{Kepler}, 
   SOPHIE and HARPS-N}           
            
   \author{
G.~H\'ebrard\inst{1,2}
\and J.-M.~Almenara \inst{3}
\and A.~Santerne \inst{3,4} 
\and M.~Deleuil \inst{3}
\and C.~Damiani \inst{3}
\and A.\,S. Bonomo \inst{3,5}
\and F.~Bouchy \inst{3}
\and G.~Bruno \inst{3}
\and R.\,F. D\'{\i}az \inst{3}
\and G.~Montagnier \inst{1,2}
\and C.~Moutou \inst{3}
}

   \institute{
Institut d'Astrophysique de Paris, UMR7095 CNRS, Universit\'e Pierre \& Marie Curie, 
98bis boulevard Arago, 75014 Paris, France 
\email{hebrard@iap.fr}
\and
Observatoire de Haute-Provence, CNRS/OAMP, 04870 Saint-Michel-l'Observatoire, France
\and
Aix Marseille Universit\'e, CNRS, LAM (Laboratoire d'Astrophysique de Marseille) UMR 7326, 13388, Marseille, France
\and
Centro de Astrof{\'\i}sica, Universidade do Porto, Rua das Estrelas, 4150-762 Porto, Portugal
\and
INAF - Osservatorio Astrofisico di Torino, Via Osservatorio 20, 10025 Pino Torinese, Italy
}

   \date{Received TBC; accepted TBC}
      
  \abstract{
We present the detection and characterization of the two new transiting, close-in, giant 
extrasolar planets \kdb\ and \khb. They were first identified by the \kepler\ team
as promising candidates from photometry of the \kepler\ satellite, 
then we established their planetary nature thanks to the  radial velocity follow-up jointly secured 
with the spectrographs SOPHIE and HARPS-N. Combined analyses of the whole datasets 
allow the two planetary systems  to be characterized. 
The planet \kdb\ has mass and radius of $0.68 \pm 0.09$~\MJ\ and $1.32 \pm 0.14$~\RJ;
it orbits in 7.34~days a F8V host star with mass and radius of $1.40 ^{+0.14}_{-0.11}$~\Msun\ 
and $1.51  \pm 0.14$~\Rsun.
\khb\ is a massive planet with mass and radius of $9.9 \pm 0.5$~\MJ\ and $1.03 \pm 0.06$~\RJ;
it orbits in 8.88~days an active G8V star with a rotation period of $19.2\pm0.3$~days, and 
mass and radius of $0.88\pm0.06$~\Msun\ and $0.88\pm0.04$~\Rsun.
Both planets lie on eccentric orbits and are located just at the frontier between regimes where 
the tides can explain circularization and where tidal effects are negligible. 
The two planets are among the first ones detected and characterized thanks to observations 
secured with HARPS-N, the new spectrograph recently mounted at the \emph{Telescopio Nazionale Galileo}.
These results illustrate the benefits that could be obtained from joint studies using two spectrographs
as SOPHIE and~HARPS-N.
}

   \keywords{Planetary systems -- Techniques: radial velocities -- Techniques: photometric -- 
   Techniques: spectroscopic -- Stars: individual: KOI-200, KOI-889}

  \authorrunning{H\'ebrard et al.}
\titlerunning{Two new transiting exoplanets KOI-200\,b and KOI-889\,b}

   \maketitle 


\section{Introduction}

Exoplanets transiting in front of their host stars allow numerous key studies to be 
performed, including accurate radius, mass, and density measurements, 
atmospheric studies in absorption through transits and in emission through occultations, 
dynamic analyses from possible timing variations, or obliquity measurements 
(see, e.g., Winn~\cite{winn10} for a review). Today, nearly 300 transiting exoplanets have 
been discovered. They were mainly detected from ground-based photometric surveys 
which are mostly sensitive to close-in giant planets
(see, e.g., H\'ebrard et al.~\cite{hebrard13}), and from their space-based counterparts 
\corot\ and \kepler\ which are sensitive to similar planets as well as to planets on longer 
periods and/or smaller radii.
The \kepler\ satellite, in particular, is monitoring stars with high-precision optical 
photometry with the goal to detect signatures of exoplanetary transits. 
About 156\,000 stars with magnitudes $9<V<16$ are continuously observed by \kepler\ 
for more than three years now.
Several candidates lists have been successively released by 
Borucki et al.~(\cite{borucki11a}; \cite{borucki11b}) then Batalha et al.~(\cite{batalha12}) 
with a list now of 2321 \kepler\ Objects of Interest (KOIs). An~additional set of 461~new 
\kepler\ planet candidates has been recently announced by Burke et al.~(\cite{burke13}). 
That  dataset provides precious inputs for exoplanetology research, for 
statistical analyses as well as for individual studies on particular KOIs.
However in most cases photometry alone does not allow the planetary 
nature of the detected transits to be established. 
Indeed, several stellar configurations can mimic planetary transits 
(e.g.~Almenara et al.~\cite{almenara09}; 
Bouchy et al.~\cite{bouchy09a}), including undiluted eclipsing binaries with 
low-mass stellar companions or diluted eclipsing binaries, namely ``blends".
Whereas such impostors represent the majority of transiting planet candidates 
detected by ground-based surveys or even by \corot, it has been argued from 
statistical studies that they should be particularly rare for \kepler\ candidates 
(e.g.~Morton \&\ Johnson~\cite{morton11}).
Other studies have shown however that a significant part of the KOIs actually 
are not caused by planets but  by scenarios implying stellar objects only 
(e.g. Bouchy et al.~\cite{bouchy11}; Col\`on et al.~\cite{colon12}; 
Santerne et al.~\cite{santerne12}). This is particularly true for 
close-in, giant exoplanet KOI candidates, whose false positive 
rate have been estimated to $34.8\pm6.5$\%\ (Santerne et al.~\cite{santerne12})
or $29.3\pm3.1$\%\ (Fressin et al.~\cite{fressin13}).

In numerous cases radial velocities (RV) allow blend scenarios and actual 
planetary transits to be distinguished. In complement to photometric data, 
radial velocities also provide additional parameters for the identified 
planetary systems; in particular they allow the mass of the planets and the eccentricity 
of their orbits to be measured. In order to do such studies on \kepler\ candidates
we started in 2010 a radial-velocity follow-up of KOIs with the SOPHIE spectrograph 
at Observatoire de Haute-Provence (OHP, France). We mainly focussed 
on the brightest stars (\kepler\ magnitude $K_p < 14.7$) harboring candidates of 
close-in, giant planets. In particular, this already allowed us to 
identify and characterize several new transiting planets 
(Santerne et al.~\cite{santerne11a}; \cite{santerne11b}; Bonomo et al.~\cite{bonomo12})
as well as more massive companions and false positives 
(Ehrenreich et al.~\cite{ehrenreich11}; Bouchy et al.~\cite{bouchy11}; Santerne et 
al.~\cite{santerne12}; D\'{\i}az et al.~\cite{diaz13}).

The installation of the HARPS-N spectrograph at the \emph{Telescopio Nazionale Galileo}
(TNG, La Palma, Spain) gives us the opportunity to extend that on-going program. 
Indeed, HARPS-N is supposed to reach a better radial-velocity precision than SOPHIE, 
especially for faint \kepler\ targets thanks to the 3.58-m diameter of the TNG, 
to be compared with the 1.93-m diameter of the OHP telescope hosting SOPHIE.
One of the main objectives of HARPS-N is the RV follow-up of the \kepler\ candidates. 
First we used HARPS-N to follow KOIs for which our SOPHIE data suggested planet 
detection but with a precision preventing firm conclusion and accurate characterization. 
Secondly we also used HARPS-N to follow KOIs fainter than the limit $K_p = 14.7$ adopted 
with SOPHIE. 
We present here two new transiting planet detections, one in each of the two above categories.
This illustrates the benefits that could be obtained from joint studies using two spectrographs
as HARPS-N and SOPHIE for the follow-up of transiting planet candidates.
We describe the photometric and spectroscopic observations of both targets in 
Sect.~\ref{sect_observations} and the analysis of the whole datasets and the results 
in Sect.~\ref{sect_analysis}. Tidal evolution of both planets is discussed in 
Sect.~\ref{sect_tidal} and we conclude in Sect.~\ref{sect_concle}.

\begin{table}[h]
\centering
\caption{IDs, coordinates, and magnitudes of the planet-host stars.}            
\begin{tabular}{lcc}       
\hline                
 Object & KOI-200 & KOI-889 \\
\hline 
\kepler\ ID 		& 6046540 			& 757450 			\\
USNO-A2 ID  	& 1275-11623662		& 11200-11449160		\\
2MASS ID   	& 19322220+4121198  	& 19243302+3634385 	\\
\hline            
RA (J2000)   & 19:32:22.21  & 19:24:33.02  \\   
DEC (J2000) &  41:21:19.87   & 36:34:38.57 \\  
\hline
\kepler\ magn. $K_{\rm p}$	& 14.41 &  15.26 \\
GSC-$V$		&	14.23 & -- \\
SDSS-$G$	&	14.876	&	15.998 \\ 
SDSS-$R$	&	14.350 	&	15.207 \\   
SDSS-$I$	&	14.212 	&	14.949 \\     
2MASS-$J$  	&	$13.306\pm  0.024$	&	$13.665 \pm 0.021$ \\ 
2MASS-$H$ 	&	$13.008\pm  0.025$	&	$13.262\pm  0.025$ \\
2MASS-$K_s$	&	$12.958\pm  0.033$	&	$13.118 \pm 0.029$ \\
WISE-$W1$	&	$12.857 \pm 0.025$	&	$13.149 \pm 0.025$ \\     
WISE-$W2$	&	$12.912 \pm 0.029$	&	$13.286 \pm 0.035$ \\
WISE-$W3$	&	$12.307 \pm 0.215$	&	      --         \\                              
\hline
\end{tabular}
\label{startable_KOI}      
\end{table}

\section{Observations and data reduction}
\label{sect_observations}

\subsection{Photometric detection with \kepler}
\label{kepphot}

\begin{figure}[ht]
 \centering
\hspace{-0.47cm}
 \includegraphics[scale=0.37]{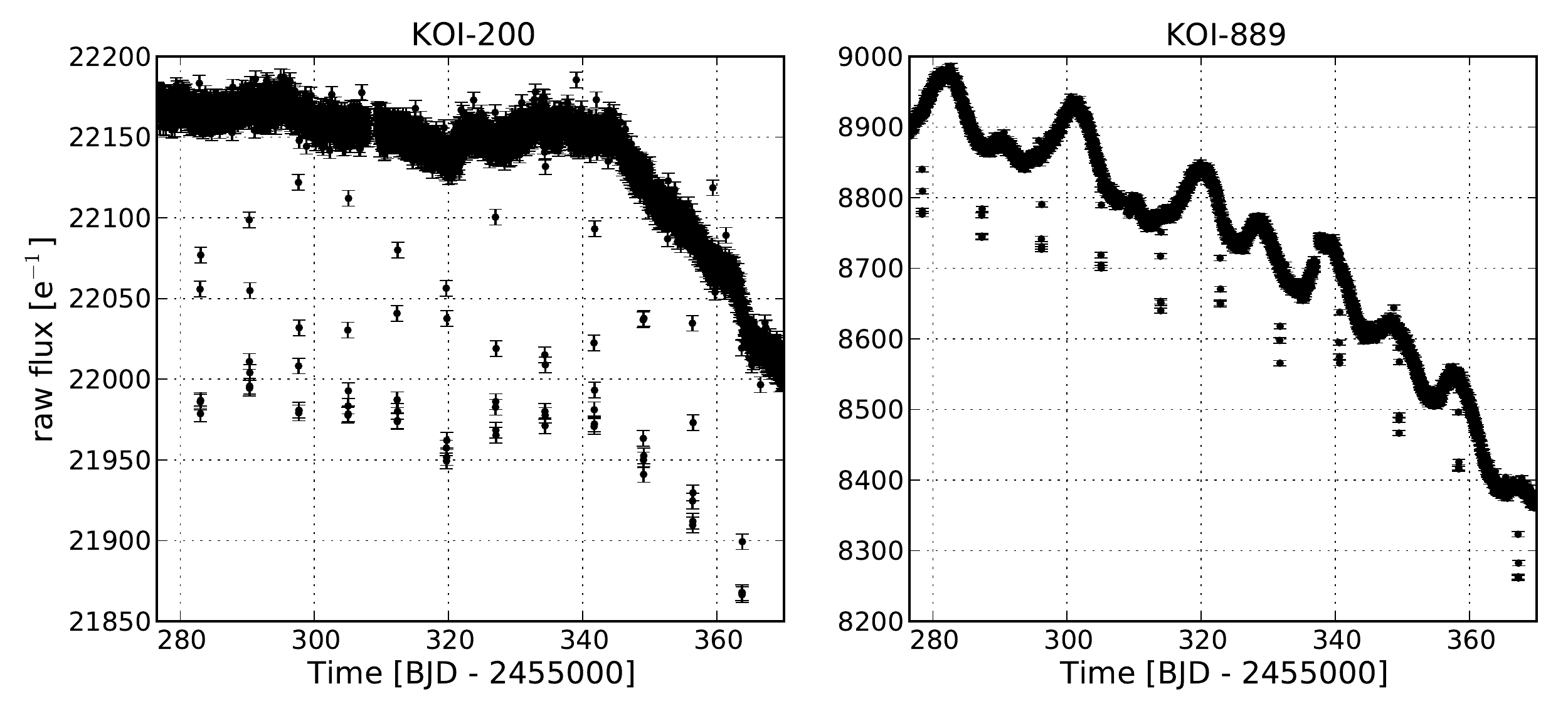}
  \caption{Time samples of the \kepler\ light curves of \kd\ and \kh\ (Q5 data only).
The signatures of periodic transits are easily detected, with periods of 
7.34 and 8.88~days respectively. In addition to the transit signal, 
the \kh\ light curve shows a modulation with a period 19.2~days, 
which is likely due to appearing spots on the rotating surface of the star.
    }
  \label{fig_photom_KOI889}
\end{figure}

The IDs, coordinates, and magnitudes of the two targets are reported 
in Table~\ref{startable_KOI}.
Both \kd\ and \kh\ were observed by \kepler\ since the beginning 
of the mission in May 2009. They were identified by Borucki et 
al.~(\cite{borucki11a}; \cite{borucki11b}) and Batalha et al.~(\cite{batalha12})
as hosting single periodic transits with periods of 7.34 and 8.88~days and 
depths characteristic of giant planets (Fig.~\ref{fig_photom_KOI889}).
No transits with different periods were detected in 
any of the light curves so there are no signs for multiple transiting systems.
The \kepler\ photometry was acquired with a time sampling of 29.4~minutes 
(long-cadence data).
The \kepler\ observations are divided in ``quarters''; at
the time of writing this paper, 
all the 13~first quarters observed 
so-far by \kepler\ were publicly available from the MAST 
archive\footnote{http://archive.stsci.edu/kepler/data\_search/search.php}. 
Some quarters were also observed in short cadence (every 1~minute) for both targets. 
Since their transits in long-cadence data are well enough sampled 
these short-cadence data were not used, which 
reduces computation time. We used the light-curve of quarters Q1 to Q13
reduced by the Photometric Analysis 
\kepler\ pipeline that accounts for barycentric, cosmic ray, background and 
``argabrightening'' corrections (Jenkins et al.~\cite{jenkins10}). Both light curves clearly present 
transits with depths of about 1\,\%\ 
as reported by Borucki et al.~(\cite{borucki11a}; \cite{borucki11b}) 
and Batalha et al.~(\cite{batalha12}), with a typical uncertainty per data point at the level of 
$\sim200$~ppm for \kd\ and $\sim500$~ppm for \kh. While the light curve of \kd\ does not show 
any sign of stellar variability, 
\kh\ shows flux modulations at levels between $\sim2$\,\%\ and $\sim4$\,\%\ peak-to-peak, 
depending of the epochs. An example is shown in the right panel of Fig.~\ref{fig_photom_KOI889}.
This is likely due to appearing spots with variable extents and locations 
on the rotating surface of the star.
Analyses through Lomb-Scargle periodogram and autocorrelation function provide the same 
values for the periodicity of the modulations, which corresponds to the rotational period 
of the star: $P_{\rm rot} = 19.2\pm0.3$~days. This is near to twice the orbital period of~the planet.
A stellar rotation period similar to the orbital period is unlikely according to the shape of the light 
curve and the~periodogram.

\begin{figure*}[]
\begin{center}
\begin{tabular}{cc}
\includegraphics[width=\columnwidth]{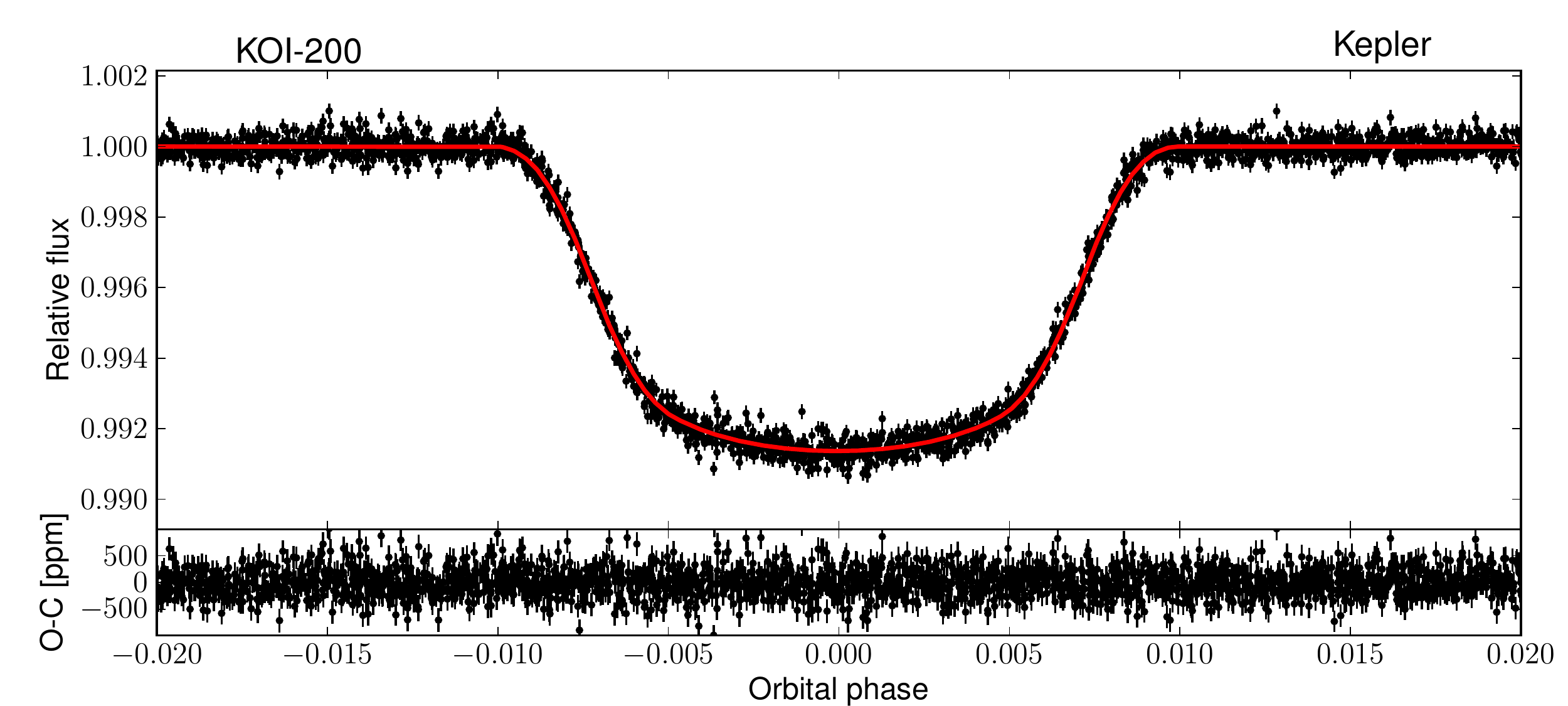} & \includegraphics[width=\columnwidth]{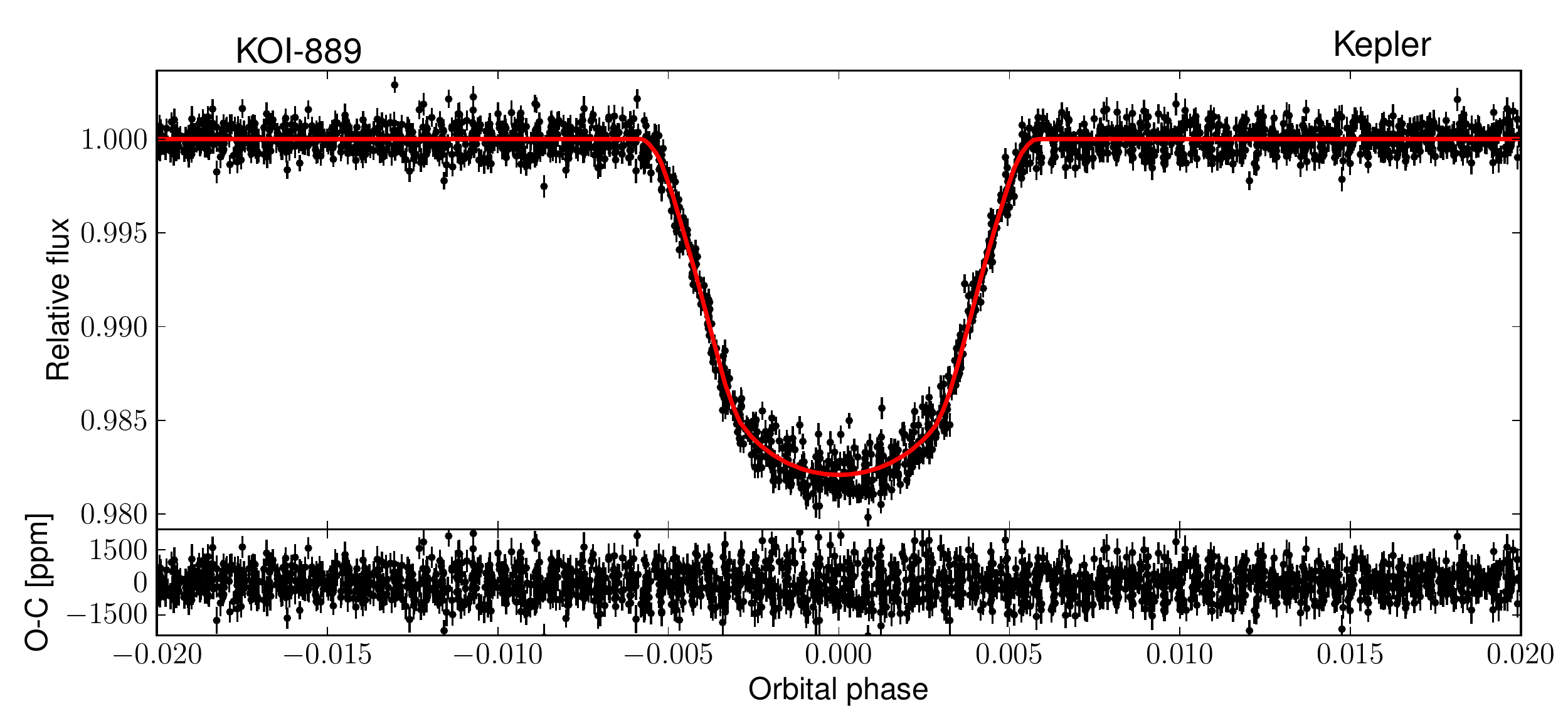}\\
\includegraphics[width=\columnwidth]{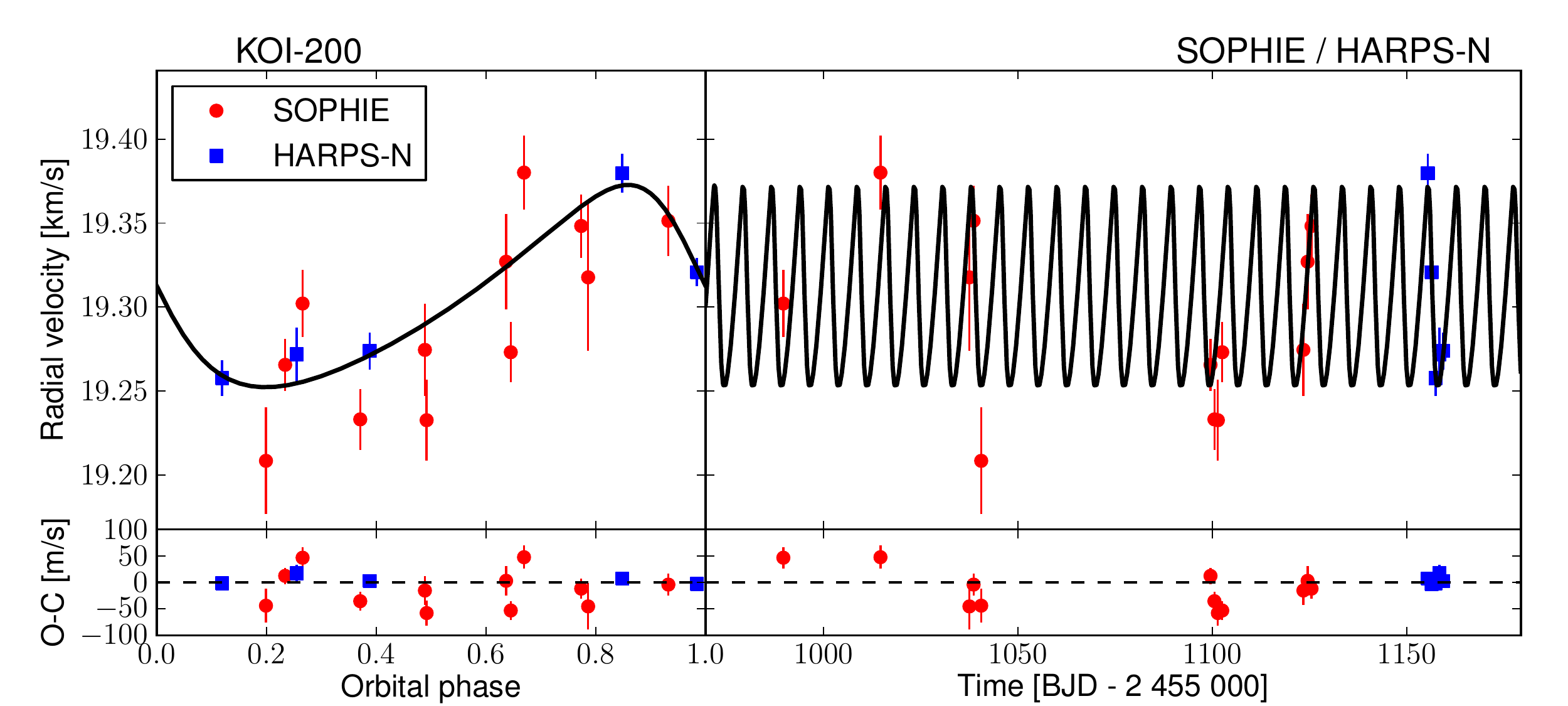} & \includegraphics[width=\columnwidth]{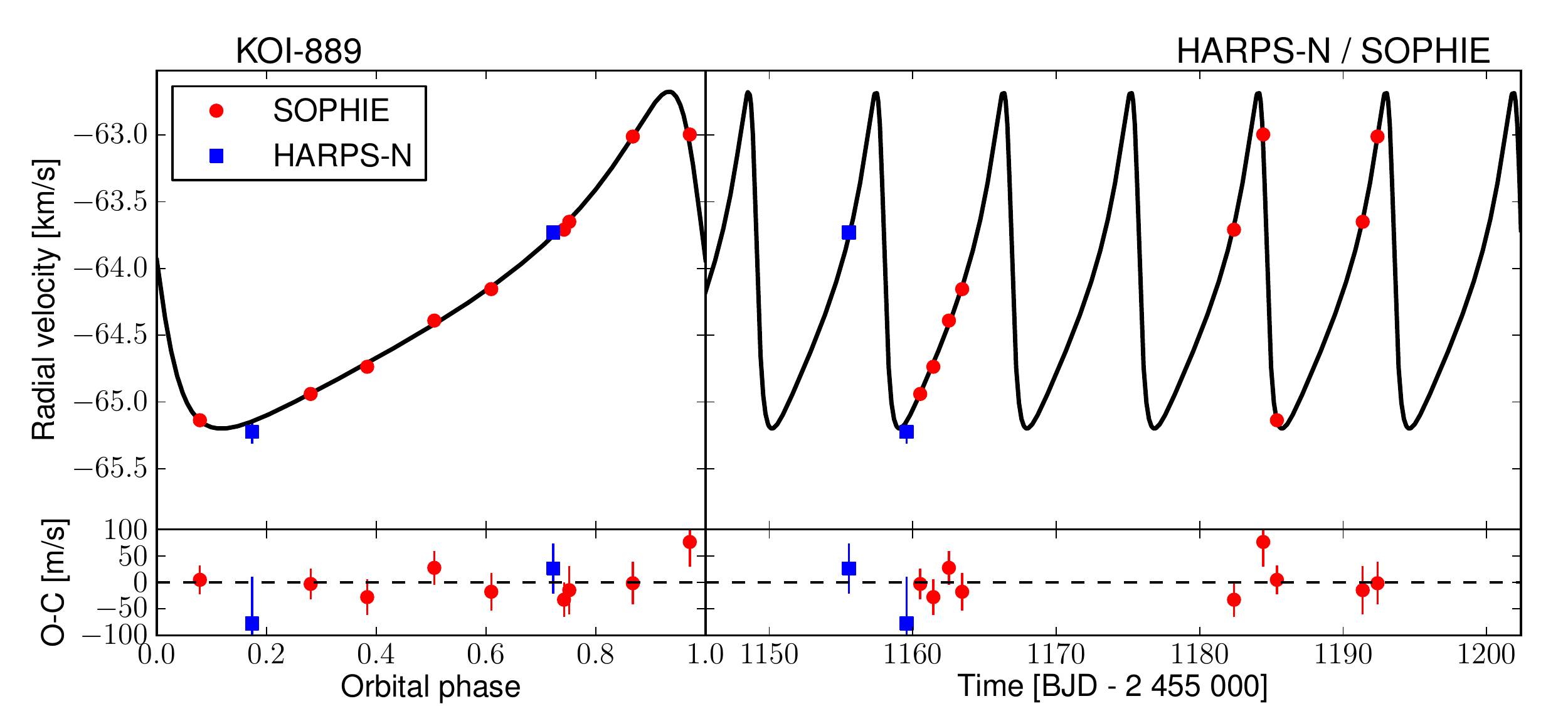}\\
\includegraphics[width=\columnwidth]{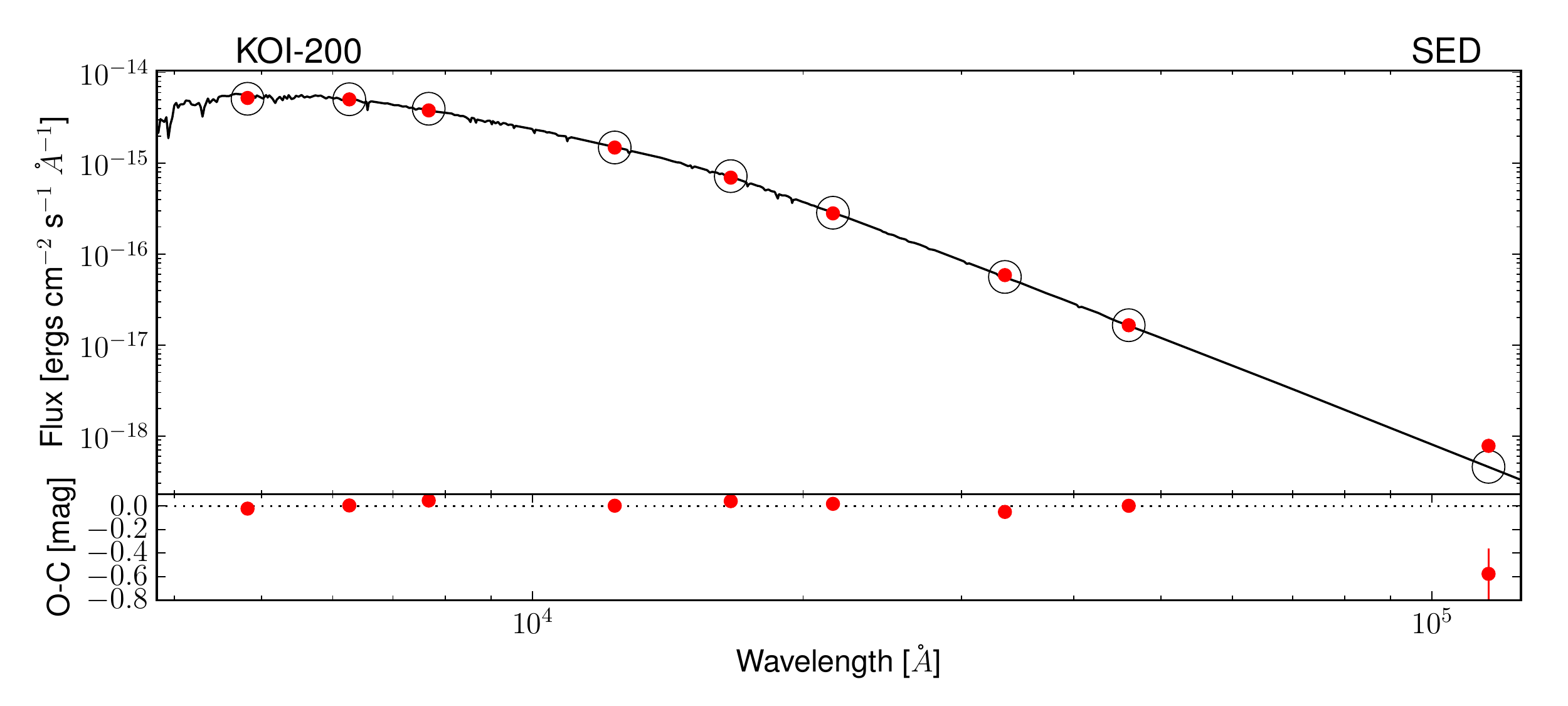} & \includegraphics[width=\columnwidth]{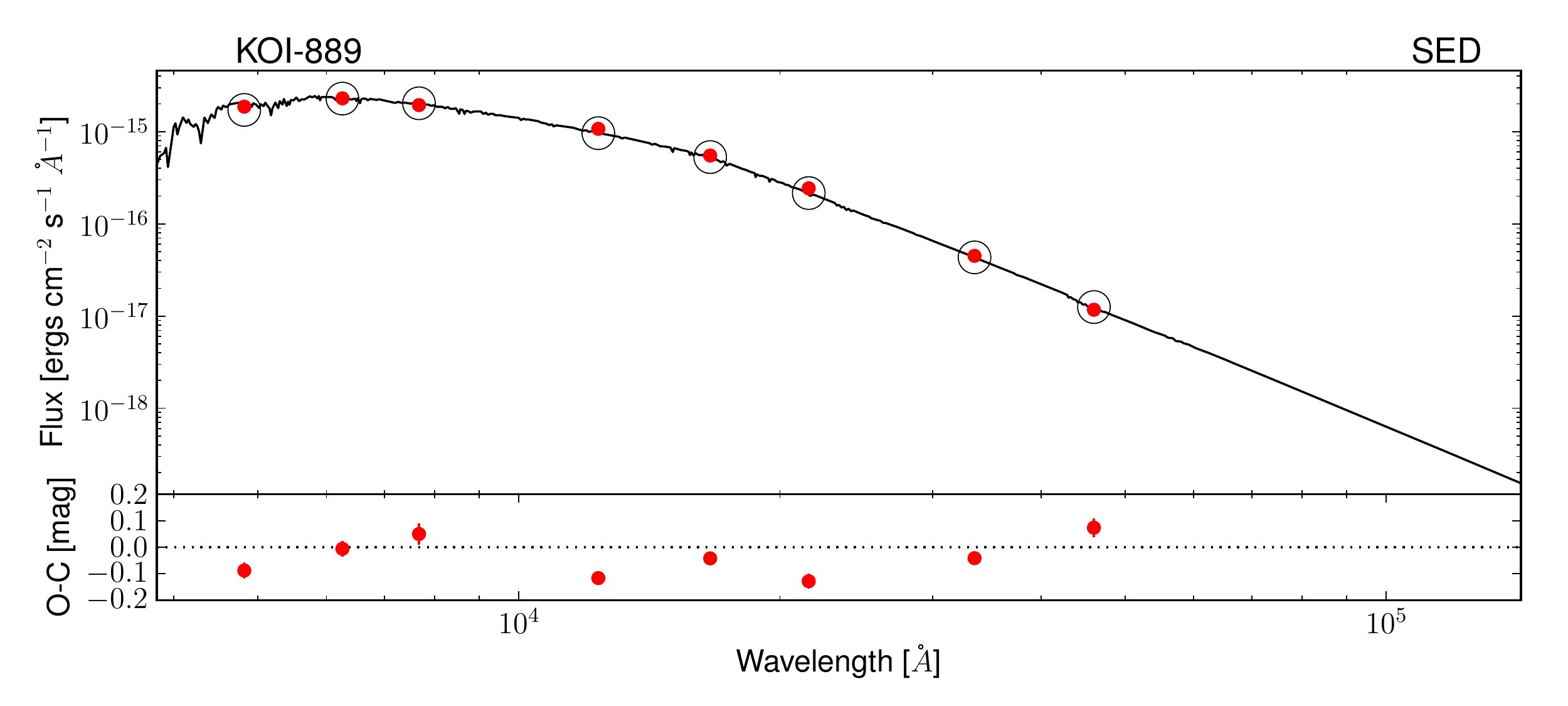} \\
\end{tabular}
\caption{
Data and best fit models for \kd\ (left) and \kh\ (right). 
The parameters of the fits are reported in Table~\ref{posterior}.
\textit{Upper panels:} \kepler\  phase-folded light-curve (black dots with 1-$\sigma$ error bars) 
over-plotted with the best model (red line), and residuals. 
\textit{Middle panels:} SOPHIE (red circles) and HARPS-N (blue squares) radial velocities and 1-$\sigma$ 
error bars phase-folded to the orbital period of the planet (on left) or as function of time (on right) over-plotted 
with the best model (black line), and residuals.
\textit{Lower panels:} Fitted spectral energy distribution (black line) over-plotted with measured magnitudes 
(red circles, Table~\ref{startable_KOI}) and corresponding integrated flux in each of the photometric bands 
according the~fit~(open~circles).}
\label{bestmodel}
\end{center}
\end{figure*}

Before modeling the transits, we normalized fragments of the light curves by fitting an 
out-of-transit parabola, 
first without accounting for contamination. Since the \kepler\  spacecraft is rotating four 
times a year, the crowding values are different between seasons. We thus produced four 
crowding-uncorrected detrended light curves for both targets, one per season. This will 
allow us to account for differential crowding values, noises and out-of-transit fluxes 
in the transit modeling and in the final 
error budget (see Sect.~\ref{sect_parameter_system}). The corresponding phase-folded 
light curves are plotted in the upper panels of~Fig.~\ref{bestmodel}.

\subsection{Radial velocities with SOPHIE and HARPS-N}
\label{sect_RV}

Both SOPHIE (Perruchot et al.~\cite{perruchot08}; Bouchy et 
al.~\cite{bouchy09b}) and HARPS-N (Cosentino et al.~\cite{cosentino12})
are cross-dispersed, stabilized echelle spectrographs dedicated to high-precision 
radial velocity measurements. Each of them is fed by a set of two optical fibers 
mounted at the focus of their telescope (1.93-m OHP for SOPHIE and 3.58-m TNG for  
 HARPS-N).  
Their spectral ranges are similar (about 385~nm~--~693~nm) and 
their wavelength calibration is made using thorium-argon lamps. 
They are the only two spectrographs fed by octagonal-section fibers
available now on the sky. This 
allows an improved stability and uniformity of the illumination for  
high-precision spectroscopy. SOPHIE was equipped by octagonal-section fibers in Spring 2011 
(Bouchy et al.~\cite{bouchy13}) and HARPS-N since its start in Spring 2012.
Both instruments are located in a thermally-controlled room but while  
only the dispersive components are contained in a sealed constant-pressure tank 
in the case of SOPHIE, the whole HARPS-N instrument is encapsulated
in a vacuum vessel which provides an even better radial-velocity stability.
For the present studies SOPHIE was used in High-efficiency mode, yielding 
a resolution power $\lambda/\Delta\lambda=40\,000$, whereas HARPS-N 
provides $\lambda/\Delta\lambda=115\,000$. 
The slow read-out modes were used for the detectors of both instruments.
The present HARPS-N observations 
were obtained just before the failure of the red side of the CCD in late September 2012. 
For both stars and both instruments the two optical-fiber 
apertures (3''- and 1''-wide for SOPHIE and HARPS-N, respectively) were used.
The first aperture was centered on the target and the second one on the sky 
to simultaneously measure its background. It allowed us to confirm there was 
no Moonlight pollution in any of our spectra significantly altering the 
radial velocity measurement.
In the case of \kh\ another star was coincidentally present in the second SOPHIE aperture. 
That second star does not affect the spectra extraction procedure 
for the first aperture, and we
double checked there was no sky pollution using the 
observations of other targets secured the nights we observed~\kh.

The spectra were extracted from the detector images with the SOPHIE and HARPS-N 
pipelines. Based on similar structures, they  include localization of the spectral 
orders on the 2D-images, optimal order extraction, cosmic-ray rejection, wavelength 
calibration, and corrections of flat-field. Then the spectra passed through weighted 
cross-correlation with G2-type numerical masks following the method described 
by Baranne et al.~(\cite{baranne96}) and Pepe et al.~(\cite{pepe02}). 
We adjusted the number of spectral orders used in the cross-correlations 
to reduce the dispersion of the measurements. Indeed, some spectral domains  
in the blue part of the spectra have particularly low signal-to-noise ratios (SNR), 
so using them degrades the 
precision of the radial-velocity measurement. At the end, we did not used the 
15 first blue orders of the 70 available ones in the HARPS-N spectra 
for the cross-correlation, as well as the 
13 and 19 first ones of the 39 SOPHIE orders for \kd\ and \kh,~respectively. 

\begin{figure}[]
 \centering
 \includegraphics[scale=0.37]{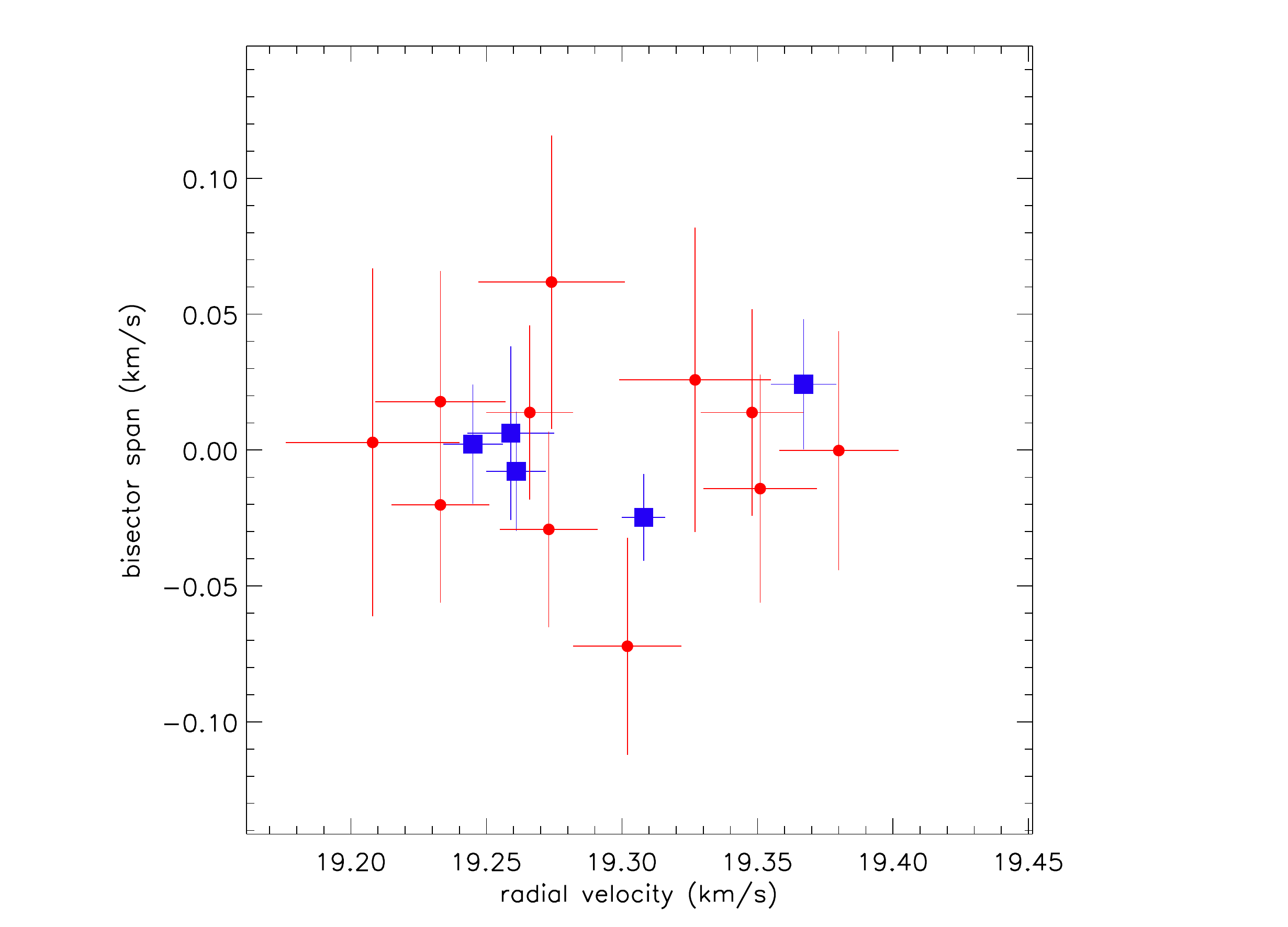}
 \includegraphics[scale=0.37]{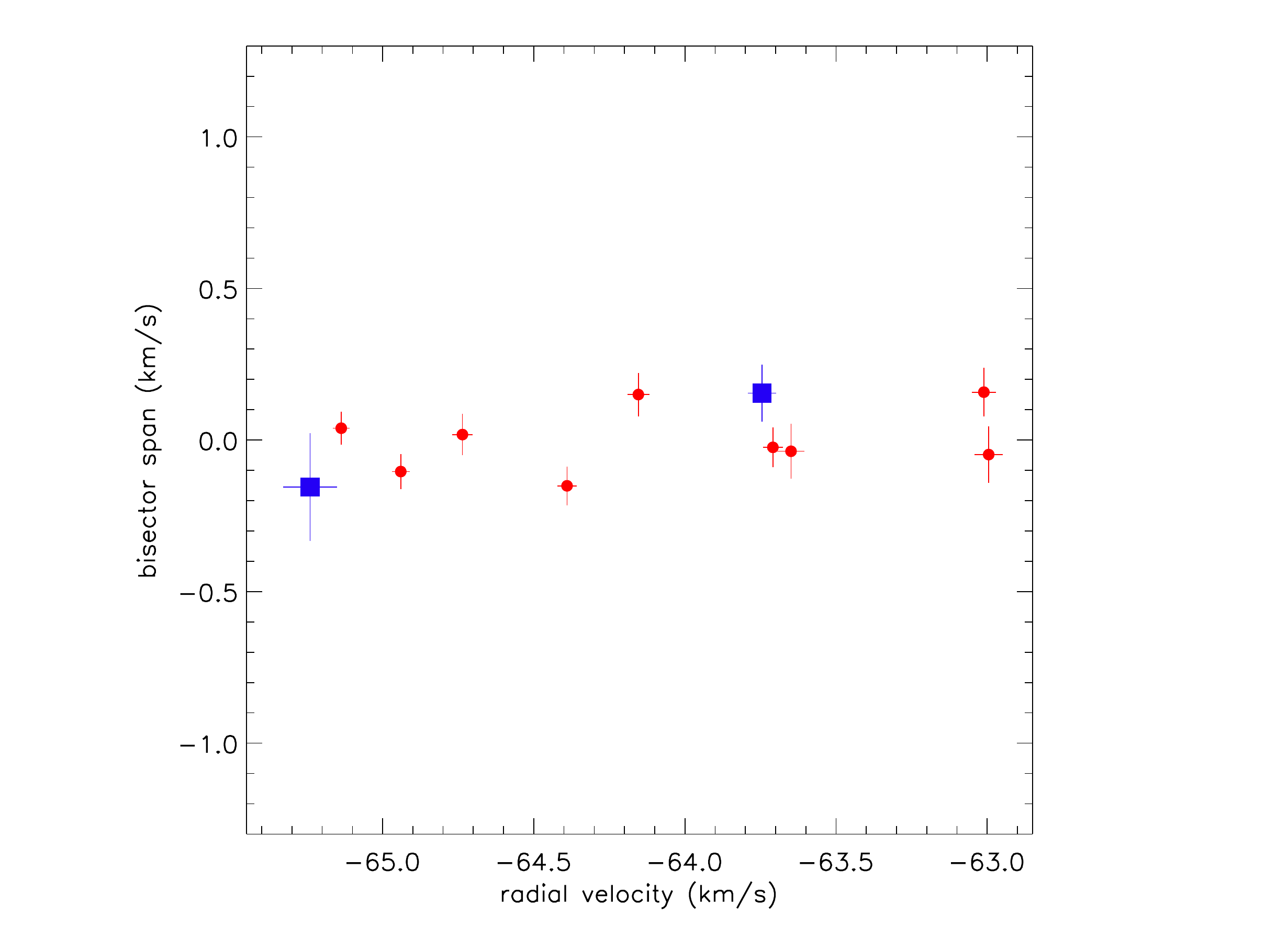}
  \caption{Bisector span as a function of the radial velocities with 1-$\sigma$\,Êerror bars
for \kd\ (top) and \kh\ (bottom). 
SOPHIE data are red circles and HARPS-N are blue squares.
The ranges have the same extents in the $x$- and $y$-axes
on each figure.
    }
  \label{fig_biss}
\end{figure}

All the exposures provide a well-defined, single peak in the
cross-correlation function (CCF).
For \kd\ their full widths at half maximum are $10.88 \pm 0.12$~\kms\ 
and their contrasts represent $23\pm4$~\%\  of the continuum for SOPHIE, 
the corresponding values for HARPS-N being
$8.38 \pm 0.06$~\kms\ and $42.3\pm0.7$~\%.
In the case of \kh\ the  values are $10.17 \pm 0.10$~\kms\ 
and $21\pm4$~\%\ for SOPHIE, and $7.49 \pm 0.14$~\kms\ 
and $54.7\pm1.2$~\%\ for HARPS-N. The differences are mainly 
due to spectral resolutions of both instruments. 
The radial velocities were obtained from Gaussian fits of the CCFs, 
together with their associated error bars and bisector spans.
They were also corrected from the interpolated drift of the spectrographs.
The bisector spans error bars were estimated to be two times those 
of the corresponding radial velocities. 
The measurements are reported in Table~\ref{table_rv} and 
plotted in Fig.~\ref{bestmodel} (middle panel) and Fig.~\ref{fig_biss}.

\begin{table}
  \caption{SOPHIE and HARPS-N measurements for the planet-host stars \kd\ and \kh.}
\begin{tabular}{cccrrrr}
\hline
BJD$_{\rm UTC}$ & RV & $\pm$$1\,\sigma$ & bisect.$^\ast$ & exp. & SNR$^\dagger$ & instr.$^\ddagger$\\
-2\,450\,000 & (km/s) & (km/s) & (km/s)  & (sec) &  \\
\hline
\multicolumn{3}{l}{\hspace{-0.2cm}\emph{KOI-200:}}  \\
5989.6552  		&	19.302		&	0.020	&	-0.106	&  2700	&	17.7 &  SOP\\ 
6014.6390  		&	19.380		&	0.022	&	-0.034	&  2490	&	17.0 &  SOP\\
6037.5189  		&	19.318		&	0.044	&	-0.206	&  3600	&	12.4 &  SOP\\
6038.5923  		&	19.351		&	0.021	&	-0.048	&  3600	&	17.7 &  SOP\\
6040.5504  		&	19.208		&	0.032	&	-0.031	&  3600	&	14.8 &  SOP\\
6099.5331  		&	19.266		&	0.016	&	-0.020	&  3600	&	19.3 &  SOP\\
6100.5370  		&	19.233		&	0.018	&	-0.054	&  3600	&	19.6 &  SOP\\
6101.4251  		&	19.233		&	0.024	&	-0.016	&  3600	&	16.1 &  SOP\\
6102.5506  		&	19.273		&	0.018	&	-0.063	&  3600	&	19.7 &  SOP\\
6123.4243  		&	19.274		&	0.027	&	0.028	&  3600	&	13.5 &  SOP\\
6124.5103  		&	19.327		&	0.028	&	-0.008	&  3600	&	15.4 &  SOP\\
6125.5146  		&	19.348		&	0.019	&	-0.020	&  3600	&	16.9 &  SOP\\
6155.4268  		&	19.430		&	0.012	&	0.051	&  2700	&	10.0 &  HAR\\ 
6156.4254  		&	19.371		&	0.008	&	0.002	&  2700	&	15.2 &  HAR\\
6157.4150  		&	19.308		&	0.011	&	0.029	&  2700	&	13.7 &  HAR\\
6158.4134  		&	19.322		&	0.016	&	0.033	&  2700	&	  8.6 &  HAR\\
6159.3891  		&	19.324		&	0.011	&	0.019	&  2700	&	12.0 &  HAR\\
\hline
\multicolumn{3}{l}{\hspace{-0.2cm}\emph{KOI-889:}}  \\
6155.5647 		&	-63.682		&	0.047	&	0.052	&  1400	&	 2.3 &  HAR\\
6159.5726  		&	-65.177		&	0.089 	&	-0.258	&   700	&	 1.1 &  HAR\\
6160.5219  		&	-64.940		&	0.029	&	-0.155	&  3600	&	12.2 &  SOP\\
6161.4375  		&	-64.736		&	0.034	&	-0.033	&  3600	&	11.1 &  SOP\\
6162.5235  		&	-64.390		&	0.032	&	-0.202	&  3600	&	10.1 &  SOP\\
6163.4458  		&	-64.154		&	0.036	&	0.099	&  2716	&	  9.4 &  SOP\\
6182.3935  		&	-63.709		&	0.033	&	-0.075	&  3600	&	13.4 &  SOP\\
6184.4300  		&	-62.995		&	0.047	&	-0.099	&  3600	&	  7.1 &  SOP\\
6185.3837  		&	-65.137		&	0.027	&	-0.012	&  3600	&	12.3 &  SOP\\
6191.3641  		&	-63.649		&	0.045	&	-0.088	&  3600	&	11.8 &  SOP\\
6192.3927  		&	-63.011		&	0.040	&	0.107	&  3600	&	12.2 &  SOP\\
\hline
\multicolumn{7}{l}{$\ast$: bisector spans; associated error bars are twice those of RVs.} \\ 
\multicolumn{7}{l}{$\dagger$: signal-to-noise ratio per pixel at 550~nm.} \\
\multicolumn{7}{l}{$\ddagger$: instrument used: SOP for SOPHIE, HAR for HARPS-N.} \\
\label{table_rv}
\end{tabular}
\end{table}

In the case of \kd, we first obtained five SOPHIE exposures 
of $\sim1$~hour each in 
March-April 2012 that showed marginal radial velocity variations, possibly 
in agreement with the \kepler\ ephemeris. The hint of detection was 
reinforced 
with seven extra SOPHIE measurements 
secured in June-July 2012. However the detection was at the limit of the 
SOPHIE capabilities with a semi-amplitude of the order of 50~m/s whereas 
the SOPHIE measurements of \kd\ typically have a $\pm25$-m/s precision.
So we re-observed \kd\ with HARPS-N five nights in a row in 
August 2012, obtaining five 45-min exposures with better precisions, 
between $\pm8$ and $\pm16$~m/s. 
Reasonably well sampling the orbital phases of the 7.34-day 
period, the HARPS-N data clearly confirmed the planet detection and allow 
a small eccentricity to be detected (Fig.~\ref{bestmodel}, left, middle~panel).

Concerning \kh, the target was fainter than our adopted magnitude limit 
on SOPHIE so we first secured in August 2012 two observations with 
HARPS-N, near the quadratures according to the \kepler\ ephemeris. 
Due to technical issues and poor weather conditions, these exposures 
were short (23 and 12 minutes only) and provided poor signal-to-noise 
ratios, which limited the radial velocity 
precisions to $\pm47$\,m/s and $\pm89$\,m/s respectively. However, it was enough to 
detect the 1.5-\kms\ variation between the two epochs. Such a large 
variation does not necessarily require HARPS-N but 
is enough to be detected with SOPHIE despite the faintness of the target.
So starting the following night 
we switched that target to SOPHIE and finally secured nine observations 
of \kh\ in August-September 2012 at OHP, sampling the different orbital 
phases of the 8.88-day period. Thanks to better weather 
conditions and longer exposure times ($\sim1$~hour per exposure), 
SOPHIE allowed an improved radial-velocity precision to be reached~on 
that target (typically $\pm35$~m/s) by comparison with~HARPS-N.

For both targets, the joint SOPHIE and HARPS-N datasets show 
radial velocity variations in agreement with the \kepler\ ephemeris 
(Fig.~\ref{bestmodel}, middle~panel). They imply semi-amplitudes 
in two different regimes, $K\simeq60$~m/s and $K\simeq1300$~m/s for \kd\ and \kh\ 
respectively, corresponding to $\sim0.7$~\Mjup\  and $\sim10$~\Mjup\ for both 
companions: they lie in the planet-mass range. 
Radial velocities obtained using different stellar masks (F0, G2, or~K5) 
produce variations with similar amplitudes, so there is no evidence that 
the variations could be explained by blend scenarios caused by stars 
of different spectral types. Similarly, the cross-correlation function bisector
spans show neither variations nor trends as a function of radial velocity 
(Fig.~\ref{fig_biss}).
A weak correlation could be seen in the case of \kh\ but it mainly stands   
on the two low-SNR HARPS-N measurements. It disappears when only the nine higher-SNR 
SOPHIE  measurements are considered.
This reinforces~the~conclusion that the radial-velocity variations are 
not~caused by spectral-line profile changes attributable to~blends~or stellar 
activity. We thus conclude that both targets harbor transiting giant planets, 
which we hereafter designate as \kdb\ and~\khb.

\section{System characterization}
\label{sect_analysis}

\subsection{Parameters of the host stars}
\label{sect_stell_analys}

The spectral analysis of both host stars was performed with the co-added 
individual spectra obtained with SOPHIE once reduced and set on the 
rest frame. 
In the case of \kd, the second aperture which is located on the sky was 
subtracted from each spectrum in order to correct from the residual 
background~sky light. For \kh\ this was not possible because of the other star 
coincidentally located in the second aperture, which implied a higher uncertainty here.
The total signal-to-noise ratios of the co-added spectra are 75 and 64 
per resolution element~at 550~nm 
in the continuum, for \kd\ and \kh~respectively. 

We performed the spectral analysis
using the iterative spectral synthesis package VWA. As described in 
details by Bruntt et al.~(\cite{bruntt10}) and references therein, 
the atmospheric parameters \teff, \logg, and \met\ were derived from a set of 49 
\ion{Fe}{i} and nine \ion{Fe}{ii} weak lines carefully selected. 
Ionization and excitation equilibrium was imposed as well as a zero slope between 
the abundances given by individual lines and their equivalent width. 
The SME package 
(version 2.1: Valenti \& Piskunov~\cite{valenti96}; Valenti \& Fischer~\cite{valenti05}) 
was also applied as a check to both stars and provided similar results 
to those obtained with VWA, but with slightly larger error bars. 
As an additional verification, we also derived the surface gravity 
from the \ion{Ca}{i} pressure-sensitive line at 612.2~nm. 
The projected rotational velocity \vsini\ and the macroturbulence were determined 
independently from a set of isolated spectral lines in the case of \kd. 
The \vsini\ measurement agrees with that obtained 
from the width of the cross-correlation function (Sect.~\ref{sect_RV}) following the 
method presented by Boisse et al.~(\cite{boisse10}). 
Due to the poor signal-to-noise ratio of the spectra, the \vsini\  of \kh\ was 
only measured from the cross-correlation function.

The fundamental parameters of the two host stars (mass, radius, and age) were  
estimated from the comparison of the location of the star in the H-R diagram with
StarEvol evolution tracks (Lagarde et al.~\cite{lagarde12}; Palacios, private com.). 
We used the atmospheric parameters \teff\ and \met\ with their associated uncertainties  
and the distribution of stellar density derived from the transit model 
(see Sect.~\ref{sect_method} below) in the 
\kid\ minimization as described by Santerne et al.~(\cite{santerne11a}) 
to derive the stellar masses, radii and gravities.

The evolutionary tracks provide an age estimate of $2.9^{_{+1.5}}_{^{-0.8}}$~Gyr for \kd.
We found two distinct  sets of solutions for \kh: one corresponds to a young star of  
200~Myr and the second one to a more evolved main-sequence star with an age of 
$10.2\pm2.1$~Gyr. The spectra of the star do not display any sign of chromospheric 
activity in the \ion{Ca}{ii} lines and
the rotation period (Sect.~\ref{kepphot}) 
 provides gyrochronological ages of $1.7\pm0.4$~Gyr and $3.1\pm0.7$~Gyr using the relations 
of Barnes~(\cite{barnes07}) and Lanza~(\cite{lanza10}), respectively.
We thus favor the oldest solutions and finally adopt  $6\pm3$~Gyr for the age of  \kh.

The parameters of both stars derived from these procedures are given in Table~\ref{posterior}. 
\kd\ is a F8V star and \kh\ a~G8V star. Both are slowly rotating.
Our derived parameters are in  good agreement with those estimated from photometry in the 
\kepler\ Input Catalog (KIC, Brown et al.~\cite{brown11}) 
but our \teff, masses, and radii are slightly~larger.

\begin{table}[t]
\caption{List of free parameters and their priors used to model the data.}
\begin{minipage}{9cm} 
\renewcommand{\footnoterule}{}                          
\begin{center}
\begin{tabular}{lcc}
\hline
Parameter & \multicolumn{2}{c}{Prior (values)$^{\dag}$} \\
 & KOI-200 & KOI-889\\
\hline
\multicolumn{3}{l}{\hspace{-0.2cm}\emph{\kepler\ photometry:}} \\
season 0 out-of-transit flux &   \multicolumn{2}{c}{$\mathcal{N}(1, 1.10^{-4})$}\\
season 1 out-of-transit flux &   \multicolumn{2}{c}{$\mathcal{N}(1, 1.10^{-4})$}\\
season 2 out-of-transit flux &   \multicolumn{2}{c}{$\mathcal{N}(1, 1.10^{-4})$}\\
season 3 out-of-transit flux &   \multicolumn{2}{c}{$\mathcal{N}(1, 1.10^{-4})$}\\
season 0 contamination [\%] &   $\mathcal{N}(4.5, 0.5)$ & $\mathcal{N}(10.4, 0.5)$\\   
season 1 contamination [\%] &   $\mathcal{N}(3.5, 0.5)$ & $\mathcal{N}(13.8, 0.5)$\\
season 2 contamination [\%] &   $\mathcal{N}(7.2, 0.5)$ & $\mathcal{N}(16.0, 0.5)$\\
season 3 contamination [\%] &   $\mathcal{N}(4.7, 0.5)$ & $\mathcal{N}(8.2, 0.5)$\\
season 0 jitter &   \multicolumn{2}{c}{$\mathcal{J}(1.10^{-6}, 1.10^{-2})$}\\
season 1 jitter &   \multicolumn{2}{c}{$\mathcal{J}(1.10^{-6}, 1.10^{-2})$}\\
season 2 jitter &   \multicolumn{2}{c}{$\mathcal{J}(1.10^{-6}, 1.10^{-2})$}\\
season 3 jitter &   \multicolumn{2}{c}{$\mathcal{J}(1.10^{-6}, 1.10^{-2})$}\\
\hline
\multicolumn{3}{l}{\hspace{-0.2cm}\emph{Radial velocities:}} \\
SOPHIE jitter [\kms] &   \multicolumn{2}{c}{$\mathcal{J}(1.10^{-5}, 1)$}\\
HARPS-N jitter [\kms] &   \multicolumn{2}{c}{$\mathcal{J}(1.10^{-5}, 1)$}\\
Systemic RV [\kms] &   \multicolumn{2}{c}{$\mathcal{U}(-100, 100)$}\\
Instruments offset [\kms] &   \multicolumn{2}{c}{$\mathcal{N}(0, 2)$}\\
\hline
\multicolumn{3}{l}{\hspace{-0.2cm}\emph{Orbit and transit light curve:}} \\
Orbital period [d] &   $\mathcal{N}(7.341, 1.10^{-4})$ & $\mathcal{N}(8.885, 1.10^{-4})$\\
Tr. epoch [BJD - 2\,454\,900] &   $\mathcal{N}(67.34, 1.10^{-3})$ & $\mathcal{N}(102.99, 1.10^{-3})$\\
System scale $a/R_{\star}$ &   $\mathcal{J}(6, 25)$ & $\mathcal{J}(5, 70)$\\
Radius ratio $R_{\rm p}/R_{\star}$ &   \multicolumn{2}{c}{$\mathcal{J}(0.01, 0.2)$}\\
Orbital inclination [$^{\circ}$] &   \multicolumn{2}{c}{$\mathcal{U}(70, 90)$}\\
Orbital eccentricity &   $\mathcal{N}(0.29,0.09)$ & $\mathcal{U}(0, 1)$\\
Argument of periastron [$^{\circ}$] &   \multicolumn{2}{c}{$\mathcal{U}(0, 360)$}\\
Linear limb darkening  &   \multicolumn{2}{c}{$\mathcal{U}(-1.5, 1.5)$}\\
Quadratic limb darkening  &   \multicolumn{2}{c}{$\mathcal{U}(-1.5, 1.5)$}\\
RV semi-amplitude [\kms] &   $\mathcal{J}(1.10^{-3},1)$ & $\mathcal{J}(1.10^{-2}, 10)$\\
\hline
\end{tabular}
\footnotetext{$^{\dag}$ $\mathcal{N}(\mu, \sigma^{2})$: Normal distribution centered 
on $\mu$ with a width of $\sigma$; 
$\mathcal{J}(a,b)$: Jeffreys distribution between $a$ and $b$; $\mathcal{U}(a, b)$: 
Uniform distribution between $a$ 
and $b$. Prior values are discussed in Sect.~\ref{sect_method}.}
\end{center}
\end{minipage}
\label{prior}
\end{table}

\subsection{Parameters of the planetary systems}
\label{sect_parameter_system}

\subsubsection{Method}
\label{sect_method}

The four normalized \kepler\ light curves constructed from quarters Q1 to Q13 (Sect.~\ref{kepphot}) 
were fitted together with SOPHIE and HARPS-N radial velocities (Sect.~\ref{sect_RV})
using the \texttt{PASTIS} code (D\'{\i}az et al., in prep.).
That code was already used in previous analyses 
(e.g. Santerne et al.~\cite{santerne11b}; D\'{\i}az et al.~\cite{diaz13};
Kostov et al.~\cite{kostov12}). The transit light curves were modeled using the EBOP code 
(Etzel~\cite{etzel81}) extracted from the JKTEBOP package (Southworth et al.~\cite{southworth04}). 
As recommended by Kipping~(\cite{kipping10}), we used an oversampling factor 
of five when comparing the model with the light curves to account for the long integration 
time of \kepler\  light curve (Kipping \& Bakos~\cite{kipping11}). Radial velocity curves were
simultaneously fitted with eccentric Keplerian orbit. 
No significant radial velocity drifts were detected so we assume only two bodies in each Keplerian fit.
The radial velocity measurements are too sparse and the time spans are too short 
to allow useful constraints to be put on the presence of additional bodies in the systems.
For each \kepler\  light curve we 
included the out-of-transit flux and the contamination factor as free parameters. 
Contamination factors reported in the KIC (Brown et al.~\cite{brown11}) 
have been shown to be erroneous in some cases  
(see e.g. KOI-205; D\'{\i}az et al.~\cite{diaz13}) 
so we chose to fit them instead of adopting the KIC values.
We also account for additional sources of Gaussian noise in the light curves and radial velocities 
by fitting a jitter value to each data set. This is  especially appropriate for \kepler\ data 
since the star is located on different CCDs each season. 
The system and data are finally described by 26 free parameters, listed in Table~\ref{prior}.
These parameters were fitted using a Metropolis-Hasting 
Markov Chain Monte Carlo (MCMC) 
algorithm (e.g. Tegmark et al.~\cite{tegmark04}; Ford~\cite{ford06}) 
with an adaptive step size (Ford~\cite{ford06}). To better sample the posterior 
distribution in case of non-linear correlations between parameters, we applied an 
adaptive principal component analysis to the chains and jumped the parameters in 
an uncorrelated space (D\'{\i}az et al., in prep.). 
Two additional parameters were not free but  fixed in the analysis because they are negligible here
in cases of exoplanets, whereas they could have significant effects for stellar binaries.
The first one is the gravity darkening coefficient fixed to $\beta_1 = 1$
(with $\beta_1 = 4 \times \beta$; see Espinosa Lara \& Rieutord~\cite{Espinosa12}).
The second one is the mass ratio which is fixed to 0; it constraints the ellipsoid modulations
of the light curves which are not detected~here.

\begin{table*}[th]
\centering
\caption{Planetary and stellar parameters for the systems KOI-200 and KOI-889.}            
\hspace{-1.5cm}
\begin{minipage}{17cm} 
\setlength{\tabcolsep}{1cm}
\renewcommand{\footnoterule}{}                          
\begin{tabular}{lcc}        
\hline                
 & KOI-200 & KOI-889\\
\hline
\multicolumn{3}{l}{\hspace{-0.8cm}\emph{Ephemeris and orbital parameters:}} \\
Planet orbital period $P$ [days] & $7.340718 \pm 0.000001$ & $8.884924 \pm 0.000002$  \\ 
Transit epoch $T_{0}$ [BJD - 2\,454\,900] & $67.3453 \pm 0.0003$ & $102.9910 \pm 0.0002$ \\  
Periastron epoch $T_{\rm p}$ [BJD - 2\,454\,900] & $67.04^{_{+0.21}}_{^{-0.34}}$ &  $102.841 \pm 0.011$ \\  
Orbital eccentricity $e$  & $0.287 \pm 0.062$ & $0.569 \pm 0.010$ \\
Argument of periastron $\omega$ [$^{\circ}$] & $64 \pm 21$ & $63.6 \pm 1.4$ \\
Orbital inclination $i_p$ [$^{\circ}$] & $85.55 \pm 0.96$ & $89.1^{_{+0.6}}_{^{-1.0}}$ \\  
Transit duration $T_{1-4}$ [hours] & $2.699 \pm 0.038$ & $1.872 \pm 0.025$ \\
Primary impact parameter $b_{\rm prim}$ & $0.684 \pm 0.032$ & $0.14 \pm 0.14$  \\
Secondary impact parameter $b_{\rm sec}$ & $1.11 \pm 0.23$ & $0.42^{_{+0.43}}_{^{-0.30}}$  \\
\hline
\multicolumn{3}{l}{\hspace{-0.8cm}\emph{Fitted transit-related parameters:}} \\
System scale $a/R_{\star}$ & $11.8^{_{+1.4}}_{^{-0.8}}$ & $19.6 \pm 0.6$ \\
Radius ratio $k=R_{\rm p}/R_{\star}$ & $0.090 \pm 0.002$ & $0.121 \pm 0.002$ \\
Linear limb darkening coefficient u$_{a}$ &  $0.10^{_{+0.25}}_{^{-0.17}}$ & $0.53 \pm 0.09$ \\
Quadratic limb darkening coefficient u$_{b}$ & $0.6 \pm 0.4$ & $0.13 \pm 0.26$ \\
\hline
\multicolumn{3}{l}{\hspace{-0.8cm}\emph{Fitted RV-related parameters:}} \\
Semi-amplitude $K$ [\ms] & $58 \pm 7$ & $1288 \pm 24$ \\
SOPHIE systemic radial velocity  $V_{0, \mathrm{S}}$ [\kms] & $19.293^{_{+0.008}}_{^{-0.014}}$ & $-64.235 \pm 0.012$ \\
HARPS-N systemic radial velocity  $V_{0, \mathrm{H}}$ [\kms] & $19.356 \pm 0.008$ & $-64.175 \pm 0.050$ \\
SOPHIE O-C residuals [\ms] & 35 & 17 \\
HARPS-N O-C residuals [\ms] & 6 & 50 \\
\hline
\multicolumn{3}{l}{\hspace{-0.8cm}\emph{Data-related parameters:}} \\
\kepler\ season 0 contamination [\%] & $4.5 \pm 0.3$ & $10.9 \pm 0.4$ \\
\kepler\ season 1 contamination [\%] & $4.9 \pmÊ0.3$ & $13.7 \pmÊ0.4$\\
\kepler\ season 2 contamination [\%] & $5.9Ê\pm 0.3$ & $13.6 \pm 0.4$\\
\kepler\ season 3 contamination [\%] & $4.7Ê\pmÊ0.3$ & $10.2 \pmÊ0.4$\\
\kepler\ season 0 jitter [ppm] & $89 \pm 12$ & $322Ê\pm 43$\\
\kepler\ season 1 jitter [ppm] & $118 \pm 12$ & $329 \pm 33$\\
\kepler\ season 2 jitter [ppm] & $101 \pm 16$ & $328 \pmÊ36$\\
\kepler\ season 3 jitter [ppm] & $99 \pm 14$ & $271 \pm 28$\\
SOPHIE jitter [\ms] & $28 \pmÊ13$ & $ < 8$ \\
HARPS-N jitter [\ms] & $ < 5$ & $< 17$ \\ 
\hline
\multicolumn{3}{l}{\hspace{-0.8cm}\emph{Spectroscopic parameters:}} \\
Effective temperature \teff[K] & $ 6050  \pm 110  $		& $ 5330   \pm 120   $   \\
Metallicity \met\ [dex] & $ 0.34 \pm 0.14   $			& $ -0.07   \pm 0.15   $   \\   
Stellar rotational velocity {\vsini} [\kms] & $5.0 \pm 1.0$ & $3.5 \pm 1.5$ \\   
Stellar macroturbulence $v_{\rm macro}$ [\kms] & $2.0 \pm 1.0$ & -  \\
Spectral type & F8V & G8V \\
\hline
\multicolumn{3}{l}{\hspace{-0.8cm}\emph{Photometric parameter:}} \\   
Stellar rotation period [days] & - & $19.2\pm0.3$ \\
\hline
\multicolumn{3}{l}{\hspace{-0.8cm}\emph{Stellar physical parameters from combined analysis:}} \\
\mr\ [solar units] & $0.745^{_{+0.085}}_{^{-0.051}}$ & $1.084 \pm 0.035$ \\ 
Stellar density $\rho_{\star}$ [$g\;cm^{-3}$] & $0.58^{+0.22}_{-0.11}$ & $1.79\pm0.17$ \\  
Star mass $M_\star$ [\Msun] & $ 1.40 ^{+0.14}_{-0.11}$		& $0.88\pm0.06$   \\     
Star radius $R_\star$ [\Rsun] & $ 1.51  \pm 0.14  $		& $ 0.88\pm0.04$   \\  
Stellar surface gravity \logg\ &   $4.2 \pm 0.1$ 		& $ 4.5   \pm 0.1   $   \\   
Age of the star [Gyr] & $2.9^{_{+1.5}}_{^{-0.8}}$			& $6\pm3$   \\     
Distance of the system [pc] & $1330 \pm 170$ & $1140^{_{+250}}_{^{-160}}$ \\  
Interstellar extinction $E(B-V)$ & $0.16 \pm 0.03$ & $0.22^{_{+0.04}}_{^{-0.02}}$ \\  
\hline
\multicolumn{3}{l}{\hspace{-0.8cm}\emph{Planetary physical parameters from combined analysis:}} \\
Orbital semi-major axis $a$ [AU] & $0.084 \pm 0.014$ & $0.080 \pm 0.005$  \\ 
Periastron distance $(1-e)\,a$ [AU] & $0.060\pm0.011$ & $0.034\pm0.002$ \\
Apoastron distance $(1+e)\,a$ [AU] & $0.108\pm0.019$ & $0.126\pm0.008$ \\
Planet mass $M_{\rm p}$ [\Mjup] &  $0.68 \pm 0.09$  & $9.9 \pm 0.5$ \\ 
Planet radius $R_{\rm p}$[\Rjup]  & $1.32 \pm 0.14$ & $1.03 \pm 0.06$ \\
Planet density $\rho_{\rm p}$ [$g\;cm^{-3}$] & $0.37\pm0.13$ & $11\pm2$ \\ 
Planetary equilibrium temperature at $a$-distance $T_{\mathrm{p},\,a}$ (K)	& $1250 \pm 120  $	& $850 \pm 40   $   \\              
Planetary equilibrium temperature at  periastron $T_{\mathrm{p},\,\mathrm{per}}$ (K)	& $1450 \pm 150  $	& $1300\pm60$   \\              
Planetary equilibrium temperature at  apoastron $T_{\mathrm{p},\,\mathrm{apo}}$ (K)	& $1100 \pm 110  $	&  $680 \pm 30$  \\              
\hline
\vspace{-0.5cm}
\end{tabular}
\end{minipage}
\label{posterior}   
\vspace{-0.075cm}
\end{table*}

Table~\ref{prior} also lists the priors used for this analysis. 
Erroneous tabulated limb-darkening  coefficients 
could alter the results (see e.g. Csizmadia et al.~\cite{Csizmadia13}).
In order to avoid underestimation of uncertainties, we chose to let the limb-darkening 
coefficients free to vary instead of adopting tabulated values.
Priors on the contamination factors were 
chosen according to the values provided by the KIC assuming 0.5\%\ of Gaussian uncertainty. 
Priors on orbital period and epoch of first transit were centered on the value found by 
Batalha et al.~(\cite{batalha12}), with a width about 100 times larger than the reported errors 
to avoid biased solutions. 
In general, the shapes and widths of priors were chosen to be large enough 
to limit the bias on the posterior. For the eccentricity of \kdb\ 
which is not well constrained in the SOPHIE data, we used as prior the posterior distribution of 
an independent MCMC analysis of the radial velocity alone in order to faster reach 
the convergence of the final MCMC. \kd\ and \kh\ were analyzed with 
55 and 40 chains leading to a total of $7\times10^{7}$ and $4\times10^{7}$ steps, respectively. 
Each chain was started at random points drawn from the joint prior. In both cases, all chains converged 
to the same solution. We used a modified version of Geweke (1992) 
diagnostic to determine and remove the burn-in phase of each chain. We then computed the 
correlation length of each converged sub-chains before thinning them. We finally merged the 
thinned chains that results in a total of about 1000 independent samples of the 
posterior distribution for both~targets. 

The distances of both stars were determined through another MCMC algorithm by comparing the 
magnitudes reported in Table~\ref{startable_KOI} with an interpolated grid of synthetic spectra 
from the PHOENIX/BT-Settl library (Allard et al.~\cite{allard12}). We used the \teff, \met, and 
\logg\ of the host star from the spectral analysis (see Sect.~\ref{sect_stell_analys}) with their 
respective errors as prior of the MCMC and let the distance and interstellar reddening as 
free parameters to fit the spectral energy distribution.
This consisted in adjust the stellar parameters within the prior to find the best synthetic 
spectrum that match the observed magnitude after rescaling it from the distance luminosity 
of the star and correcting from the interstellar extinction.

\subsubsection{Results}

The 68.3-\%\ confidence intervals (corresponding to 1-$\sigma$ intervals assuming 
Gaussian distributions) are listed in Table~\ref{posterior}. 
The maximum posterior models are displayed in Fig.~\ref{bestmodel}.

Our procedure and the quality of the \kepler\ data allow the contamination factors to be
directly measured. They show slight variations between the 
four \kepler\ seasons, as expected due to the four different orientations of the satellite.
Our derived values are similar to the ones reported in the KIC but here we can determine 
their uncertainties. 
The derived jitter is around 100~ppm for \kd, which is compatible with the typical value 
derived by Gilliland et al.~(\cite{gilliland11}) for \kepler\ data (D\'{\i}az et al.~\cite{diaz13}).
In the case of \kh\ we found $\sim300$~ppm; that excess of jitter could be explained by 
the activity of the star.
The jitters show some variations between seasons, probably 
linked to the different characteristics of the four different parts of the detector that~are 
used each season. 
The out-of-transit fluxes fitted for each season were found near unity with typical uncertainties 
of $\pm 8$~ppm and $\pm 25$~ppm for \kd\ and \kh, respectively. 
None of the light curves show the signature of planetary occultation at the secondary eclipse 
phase, as expected for relatively long-period planets such as these.
Occultations depths are expected here to be at most of the order of a few tens of ppm.
In addition in the case of \kdb, the secondary eclipse is likely to be un-observable from the Earth 
as the secondary impact parameter is $b_{\rm sec} = 1.11 \pm 0.23$. This is the case for 
a significant part of eccentric planets, even on close-in orbits (Santerne et al.~\cite{santerne13}).
Concerning the radial velocities, only the SOPHIE data for \kd\ need a significant jitter of $28\pm13$~m/s 
to be added. For the other cases the dispersion of the residuals around the fit agree with the expected 
error bars on radial velocities. 

From the results of the spectral analysis, we expect limb-darkening coefficients from 
Claret et al.~(\cite{claret12}) of $u_{a} = 0.34 \pm 0.02$, $u_{b} = 0.30 \pm 0.10$ for 
\kd\, and $u_{a} = 0.47 \pm 0.03$, $u_{b} = 0.22 \pm 0.02$ for \kh. 
Our fitted limb-darkening coefficients are compatible with the expected values within 
1-$\sigma$. Fixing the coefficients to the values from Claret et al.~(\cite{claret12})
does not significantly change our results.
Similarly, the \logg\ values derived from the stellar spectra agree with those derived 
from the stellar density. And finally, 
the fit of the spectral energy distribution provides the distance and the interstellar 
extinction $E(B-V)$ of each target as well as the stellar parameters \logg, \teff, and \met, which were 
found to agree with the values derived from the stellar 
analysis itself.
Thus the different analyses are coherent with respect to the different constraints
they use.

Orbiting a F8 dwarf star in $7.340718 \pm 0.000001$~days, 
\kdb\ is a giant planet with a mass of $0.68 \pm 0.09$~\MJ\ and a radius of $1.32 \pm 0.14$~\RJ.
Its density $\rho_{\rm p}=0.37\pm0.13$~g/cm$^3$ is particularly low.
Whereas the SOPHIE data alone do not allow a significant eccentricity to be detected, the 
five HARPS-N measurements exclude circularity for the orbit, with a residuals dispersion reduced 
from $\pm14$~m/s to $\pm6$~m/s between circular and eccentric orbits. The light curve 
also provides constraints on the eccentricity through the duration of the transit. In addition, 
in the case of a circular orbit the derived limb darkening coefficients were in poorer agreement 
with the ones expected from spectral analysis according Claret et al.~(\cite{claret12}). 
With all the constraints we finally derived $e=0.287 \pm 0.062$. 
The eccentric solution is $\sim75$~times more probable than the circular one.
The equilibrium temperatures of the planet derived from \teff\ and $R_\star$
and assuming an isotropic zero-albedo 
are  $T_{\mathrm{p}}=1450 \pm 150$~K, $1250 \pm 120  $~K, and $1100 \pm 110  $~K 
respectively at periastron, semi-major axis distance, and apoastron. 

The massive planet \khb\ orbits a G8 dwarf star  in $8.884924 \pm 0.000002$~days. 
Its mass is $9.9 \pm 0.5$~\MJ\ and its radius $1.03 \pm 0.06$~\RJ.
It lies below the deuterium-burning mass ($\sim13$~\MJ) which could be used to distinguish 
planets from brown dwarfs. The eccentricity of the orbit is clearly detected in the radial 
velocities. The value $e=0.569 \pm 0.010$ could be accurately determined thanks to the 
large amplitude of the radial velocity variation. 
The equilibrium temperatures of the planet assuming an isotropic zero-albedo 
are  $T_{\mathrm{p}}=1300\pm60$~K, $850 \pm 40$~K, and $680 \pm 30$~K respectively at periastron, 
semi-major axis distance, and apoastron. 
Its active host star shows the signature of evolutional spots in the light curve, 
but the signal-to-noise ratio on the spectra is too low to allow that activity to be detected 
in the radial velocity jitter nor in the core of the \ion{Ca}{ii}~lines.
The deduced rotation period is $P_{\rm rot} = 19.2\pm0.3$~days which together with 
the measured stellar radius implies a rotation velocity $V_{\rm rot}\simeq2.3$~km/s. 
This is smaller than the  value \vsini\,$=4.6\pm1.0$\,km/s
measured from the lines width. This indicates a possible 
underestimation of the latest value so
we adopt  \vsini\,$=3.5\pm1.5$\,km/s to take into account for the two evaluations.
If the deduced rotation velocity $V_{\rm rot}$ had been larger than \vsini, 
it would have been consistent with $\sin i_*<1$, implying an oblique orbit of the planet 
with respect to the equatorial plane of the star. As this is not the case, 
there are no signs for spin-orbit misalignment and probably $i_* \simeq i_p \simeq 90^{\circ}$.

For both planets 
our derived parameters agree with the ones  first measured by 
the \kepler\ team (Borucki et al.~\cite{borucki11a}; \cite{borucki11b}; Batalha et al.~\cite{batalha12}).
An exception is the system scales $a/R_{\star}$ which we found to be significantly smaller 
than the ones initially reported from the first \kepler\ light curves because of the 
unconsidered eccentricities.

\section{Tidal evolution}
\label{sect_tidal}

With their orbital periods of 7.34 and 8.88~days, \kdb\ and \khb\ are in a regime 
where a modest amount of transiting planets have been detected now. It is 
growing up with the \kepler\ candidates but most of them have shallow transits.
Only ten transiting giant planets are known today in the period range 
$6-50$~days. 
The mass and period of \kdb\ are similar to those of the three planets 
WASP-59b, CoRoT-4b, and HAT-P-17b.
Concerning the massive, giant planet \khb, it has only two nearly analogs
with orbital periods longer than typical periods of hot Jupiters, namely 
HAT-P-2b and Kepler-14b. CoRoT-14b is a fourth case of massive, transiting 
planet, but its orbital period is shorter (1.5~d).
\khb\ is one of the rare known massive planets orbiting 
a G-type star in a close-in orbit. The lack of planets in that regime was underlined 
by Bouchy et al.~(\cite{bouchy11}) who proposed that such planets would not survive 
too close to G dwarfs due to engulfment caused by tidal interactions with the 
stellar convective zone. If that scenario is correct, \khb\ does not follow it.

\begin{figure}[t]
 \centering
  \vspace{-0.7cm}
  \hspace{-0.9cm}
 \includegraphics[scale=0.54]{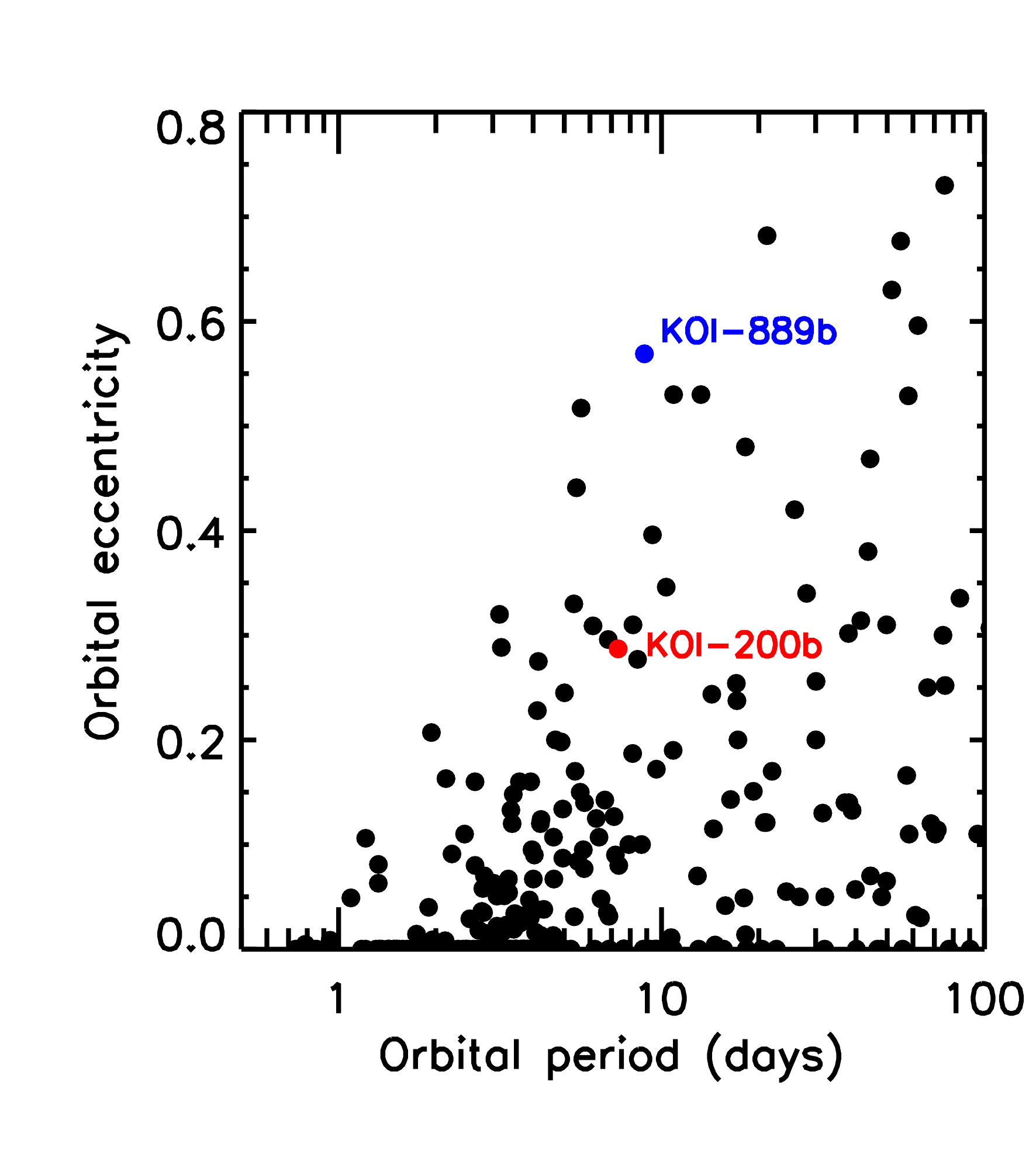}
 \vspace{-0.4cm}
  \caption{Orbital eccentricity and period of \kdb\ and \khb\ compared with other known extrasolar planets.
    }
  \label{fig_ecc}
\end{figure}

Both being on eccentric orbits, 
\kdb\ and \khb\ are part of the borderline planets 
as seen in Fig.~\ref{fig_ecc}, i.e. planets that are in the period range 
$6-30$~days 
that lie in the crossover between a regime where the tides are sufficient to explain 
circularization and a regime where tidal effects are negligible. The dispersion in 
eccentricity of these systems can be seen as the result of different spin-down rates of 
young stars, as argued by 
Dobbs-Dixon et al.~(\cite{dobbs04}). The case of \khb\ is particularly interesting since 
it is close to the envelope of the maximum observed eccentricity for systems 
in this period range.

Tidal interaction can lead either to a spiralling of the planet into the star followed by a collision, 
or to the asymptotic evolution towards an equilibrium, characterized by orbital circularity, 
co-planarity, and co-rotation. Under the assumption that the total angular momentum is conserved 
(i.e. neglecting magnetic breaking), it can be shown (Hut~\cite{hut80}) 
that equilibrium states exist only when the total angular momentum of the system $L_{\rm tot}$ 
is larger than some critical value $L_{\rm crit}$; 
such equilibrium states are unstable if the orbital angular momentum 
$h=\sqrt{G M_\star^2 M_{\rm p}^2 a (1-e^2)/(M_\star + M_{\rm p})}$ (where G is the gravitational 
constant) accounts for less than three quarters of $L_{\rm tot}$. A system that is expected to evolve 
towards a stable equilibrium is generally called ``Darwin stable". So far, the vast majority of 
known exoplanetary systems are not Darwin stable (Matsumura et al.~\cite{matsumura10}) 
and their planets are doomed to eventually fall into the host star.

In this respect, \kdb\ is not particularly remarkable despite its eccentricity. 
Using the stellar rotational velocity \vsini\  
to assess the rotational frequency $\Omega_\star$ of the star (assuming $i_*\simeq90^\circ$) 
and taking the gyration radius $\gamma$ from stellar models for the corresponding mass and 
temperature (Claret~\cite{claret95}), we can compute the stellar rotational angular momentum, 
or spin, $L_\star = M_\star (\gamma R_\star)^2 \Omega_\star$. 
This assumes that the star rotates rigidly and the interior non-uniform distribution 
of mass is accounted for by $\gamma$.  For the \kd\ system, considering the 
typical value of tidal dissipation efficiency in planets $Q'_p = 10^6$ 
(Matsumura et al.~\cite{matsumura08} ), the timescale for the synchronization and alignment 
of the planet relative to the orbit is $\sim 4\times 10^5$~yr
so we can safely assume that the planet is in pseudo-synchronization. 
Taking the gyration radius of a polytropic model of index 1 for the planet, the planetary spin 
can also be computed. Adding both spins to the orbital angular momentum $h$ gives the total 
angular momentum, which is found to be $L_{\rm tot}\simeq 0.8 \, L_{\rm crit}$. Thus the system 
is Darwin unstable. The subsequent evolution depends on the relative efficiency of the dissipation 
of the tides within the star and the planet. Either the dissipation in the planet is dominant and the 
circularization of the orbit will happen before any significant orbital decay can occur, or it is the 
efficiency of tidal dissipation in the star that will control the evolution of all the parameters, and 
the orbit will keep a non-null eccentricity until the engulfment. Finally, little can be inferred from 
the present state of the system regarding the origin of the eccentricity and the migration scenario. 
Considering the moderate value of the eccentricity, the lack of  constraints on the obliquity or the 
presence of a possible companion in the system, both planet-planet scattering or 
migration in a disk followed by weak tidal interaction are valid hypothesis.
The expected amplitude of the Rossiter-McLaughlin anomaly is $\sim30$~m/s, so the 
measurement of the obliquity through the spectroscopic observation of a transit is~feasible. 

\begin{figure}[h]
 \centering
\vspace{-1.5cm}
\hspace{-1.4cm}
 \includegraphics[scale=0.49]{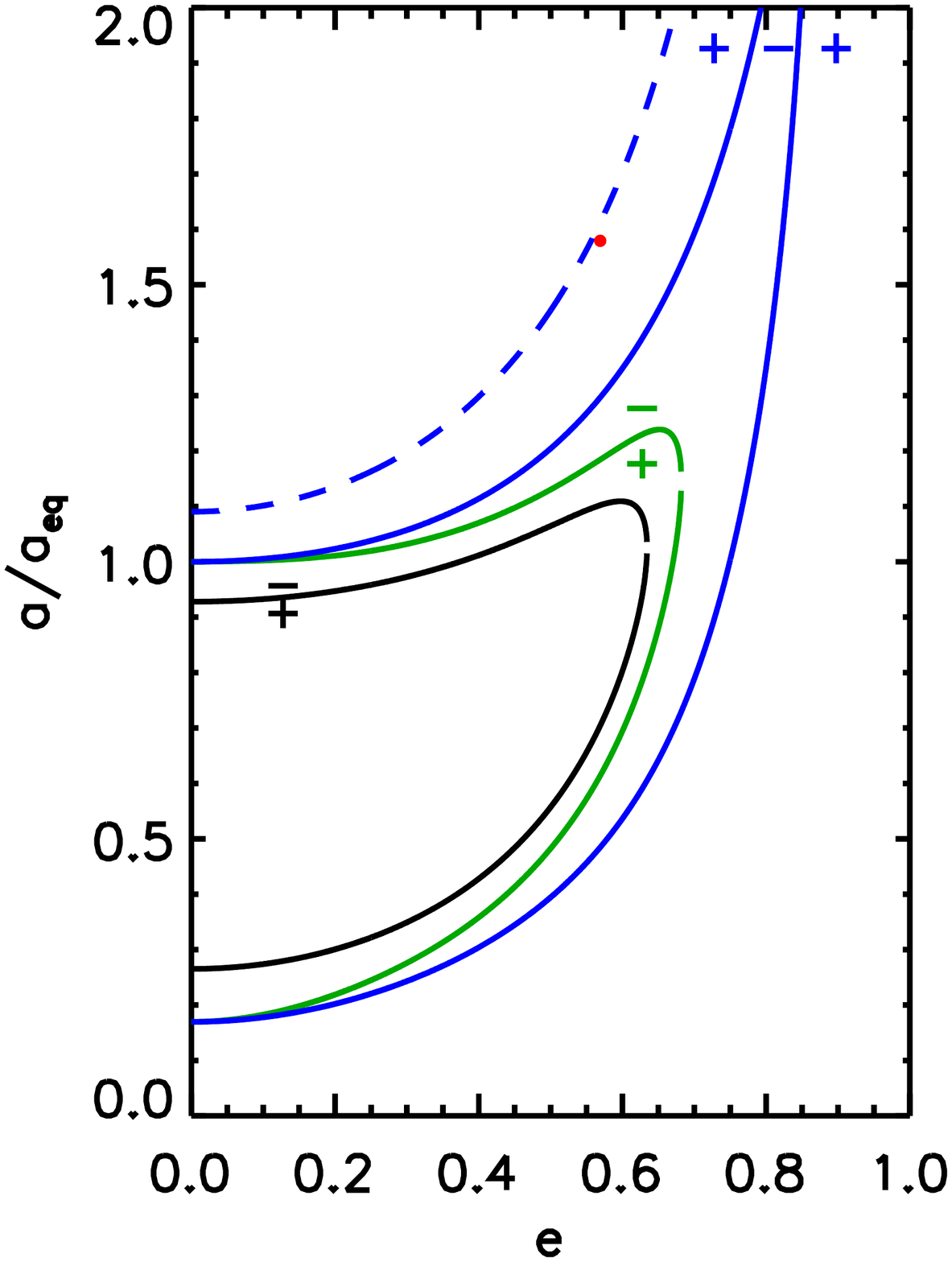}
 \vspace{-1cm}
  \caption{Regions of different types of tidal evolution for the planetary system \kh, neglecting magnetic braking 
  following Hut~(\cite{hut81}). The semi-major axis $a$ is normalized to its value at the stable 
  equilibrium $a_{\rm eq}$. The black, solid line gives the stationary points of the eccentricity, 
  with the plus and minus signs indicating the domains where the eccentricity increases and 
  decreases, respectively. The green and blue solid lines give the corresponding information 
  for the semi-major axis and rotational angular velocity of the star, respectively. The dotted 
  blue line separates the upper region where $\Omega_\star  < 0$ from the lower region 
  where $\Omega_\star  > 0$. The red dot is the present state of the system, 
  corresponding to an evolution where $d\Omega_\star/dt >0$, $da/dt < 0$ and $de/dt < 0$.
    }
  \label{fig_KOI889_tidal}
\end{figure}

On the other hand, \khb\ is much more special. The rotation period of its host star is estimated 
to $P_{\rm rot} = 19.2\pm0.3$~days (Sect.~\ref{kepphot}). Using the same reasoning 
as described above we find that the system has $L_{\rm tot}\simeq 1.3 \, L_{\rm crit}$, thus 
allowing for the existence of equilibrium states. Furthermore $h/L_{\rm tot}\simeq 0.9904$, 
meaning that the system is evolving toward the stable equilibrium. 
Considering that $n/\Omega_\star \simeq 2$ (where $n$ is the mean orbital motion), 
we know that the tides act to bring the planet closer to the star while circularizing the orbit. 
Moreover, in the frame of the weak friction model of the equilibrium tides with constant time lag, 
the main global features of tidal evolution can be derived, using energy and angular momentum 
considerations only (Hut~\cite{hut81}). 
It turns out that the only external parameter of the coupled differential equations that rule the 
temporal behavior of the orbital parameters is $\alpha$, the ratio of the orbital and rotational 
angular momentum of the system at the stable equilibrium. For \khb, $\alpha\simeq 22.8$, 
allowing for the existence of turning points in the evolution of both the semi-major axis 
and the eccentricity. The past and future evolution of the system can thus be 
non-monotonic (see Fig.~\ref{fig_KOI889_tidal}).
Considering that \kh\ is a G8V star showing clear evidence of magnetic activity, 
this problem is furthermore complicated by the effect of stellar magnetic braking, which 
is not taken into account in the derivation by Hut~(\cite{hut81}) but which 
makes the star to spin down (Dobbs-Dixon et al.~\cite{dobbs04}).

The fact that $n/\Omega_\star \simeq 2$ is intriguing because it is the value where Type~II 
migration stops if the disk is truncated at the co-rotation radius (Kuchner \& Lecar~\cite{kuchner02}).
Assuming this migration mechanism is indeed the dominant one for \khb, this would 
be an evidence that the tidal interactions in this system since the disappearance of the disk 
are either weak or that the system is young. The rotation period of the star 
invalidates the former interpretation as young stars are observed to rotate 
significantly faster, meaning that to preserve the synchronization some tidal interaction 
must have occurred. Moreover in both cases, the significant eccentricity of the orbit is 
difficult to explain because Type~II migration tends to produce circular systems 
(e.g. Ben{\'{\i}}tez-Llambay et al.~\cite{benitez11}). 
However, if we consider that the star was still in 
its pre-main-sequence phase when the disc disappeared 
(Mamajek~\cite{mamajek09}), 
the rapid increase of the spin of the star caused by the contraction of the radius 
can lead the system into a domain where the eccentricity can be excited while 
the orbit would widen, assuring the survival of the planet. The subsequent 
evolution would see the spin loss of the star caused by magnetic braking bring 
progressively the system in the domain 
where both the eccentricity and the semi-major axis are damped toward the stable equilibrium. 
However, the angular momentum extracted from the system by magnetic braking will 
eventually make the system Darwin unstable. Depending on the relative efficiency of 
tidal dissipation and magnetic braking, this might bring the planet to spiral into its host 
before the end of the main sequence lifetime of the star.

Although the age of the star \kh\ 
is not well constrained, a detailed study could give useful constraint on the efficiency of 
the tidal dissipation in both the star and planet, as well as an estimate of the loss of 
angular momentum through magnetic braking. For the massive planet \khb, a direct 
obliquity measurement would be useful to confirm or not the correlation between spin-orbit 
misalignment and planetary mass suggested by 
H\'ebrard et al.~(\cite{hebrard10}; \cite{hebrard11}).
Indeed \khb\ is one of the few known transiting planets in that range of radius and mass, 
together with HAT-P-2b, CoRoT-14b, and Kepler-14b.
Comparison between \vsini\ and $P_{\rm rot}$ does not suggest any spin-orbit misalignment
in the \kh\ system, whose expected amplitude of the Rossiter-McLaughlin anomaly 
is $\sim40$~m/s.

\section{Conclusion}
\label{sect_concle}

We have presented the detection and characterization of \kdb\ and \khb, two new transiting, close-in, 
giant extrasolar planets. They were first detected as promising candidates by the \kepler\ team from 
\kepler\ light curves.
We have established their planetary nature with the radial velocity follow-up we jointly 
secured using both the spectrographs SOPHIE and HARPS-N, and characterized them through 
combined analyses of the whole photometric and spectroscopic datasets.
The planet \kdb\ orbits its F8V host star in 7.34~days and its mass and radius are 
$0.68 \pm 0.09$~\MJ\ and $1.32 \pm 0.14$~\RJ.
\khb\ is a massive planet orbiting in 8.88~days an active G8V star with a rotation period of $19.2\pm0.3$~days; 
its mass and radius are $9.9 \pm 0.5$~\MJ\ and $1.03 \pm 0.06$~\RJ.
Both planets lie on eccentric orbits and are located just at the frontier between regimes where the tides  
can explain circularization and where tidal effects are negligible, making them interesting systems 
with respect to tidal evolution. 
\kd\ is yet another example to be already Darwin unstable whereas 
\kh\ is one of the few known currently Darwin-stable exoplanetary systems.

Since its installation in Spring 2012 
at the \emph{Telescopio Nazionale Galileo}, HARPS-N allowed in particular 
the obliquity of the 
transiting planet Qatar-1\,b to be measured (Covino et al.~\cite{covino13}) and 
the metal-poor star HIP\,11952 to be shown not to harbor giant planets 
(Desidera et al.~\cite{desidera13}).
The two new planets \kdb\ and \khb\ presented here are among the first ones to be detected
and characterized with HARPS-N. 
Our observing program managed jointly with SOPHIE and HARPS-N shows the 
benefits that could be obtained for the follow-up of transiting planet candidates
from coordinated observations secured with two spectrographs with different 
sensitivities, precisions, and accessibilities.

\begin{acknowledgements}
This publication is based on observations collected with the NASA's satellite \kepler, 
the SOPHIE spectrograph on the 1.93-m telescope at \emph{Observatoire de Haute-Provence} (CNRS), 
France (program 12A.PNP.MOUT), and 
the HARPS-N spectrograph on the 3.58-m \emph{Telescopio Nazionale Galileo}, 
La Palma (program OPT12B\_13 from OPTICON common time allocation process for EC 
supported trans-national access to European telescopes).
The authors particularly thank the \kepler, OHP, and TNG teams, whose work and expertise allow
these results to be obtained.
This publication also makes use of data products from 2MASS, 
which is a joint project of the University of Massachusetts and the Infrared Processing 
and Analysis Center/California Institute of Technology, funded by the NASA and the NSF, 
as well as data products from WISE, 
which is a joint project of the University of California and the JPL/MIT, funded by the NASA.
Funding for SDSS-III has been provided by the Alfred P. Sloan Foundation, the Participating 
Institutions, the NSF, and the U.S. Department of Energy Office of Science. 
The research leading to these results has received funding from the 
``Programme National de Plan\'etologie'' (PNP) of CNRS/INSU, 
and from the
European Community's Seventh 
Framework Programme (FP7/2007-2013) under grant agreement number RG226604 (OPTICON).
AS acknowledge the support by the European Research Council/European Community under the 
FP7 through Starting Grant agreement number 239953 and the support from Funda\c{c}\~ao para 
a Ci\^encia e a Tecnologia (FCT) in the form of grant reference PTDC/CTE-AST/098528/2008.
RFD is supported by CNES.
\end{acknowledgements}

\end{document}